\documentclass[conference]{IEEEtran}
\IEEEoverridecommandlockouts

\usepackage{amsmath,amssymb,amsfonts}
\usepackage{algorithmic}
\usepackage{graphicx}
\usepackage{textcomp}
\usepackage{xcolor}
\usepackage{amsmath}
\usepackage{hyperref}
\usepackage[sort,compress]{cite}
\usepackage{enumitem}
\usepackage{mathrsfs}
\usepackage{bbm}
\usepackage{graphicx}
\usepackage{subfigure}
\usepackage{mathrsfs}
\usepackage{multirow}
\usepackage{balance} 
\usepackage[vlined,boxed,commentsnumbered, ruled, linesnumbered]{algorithm2e}
\usepackage{amsfonts,amssymb}
\usepackage{array}
\usepackage{setspace} 
\usepackage{xcolor}
\usepackage{soul}
\usepackage{booktabs}
\usepackage[framemethod=TikZ]{mdframed}
\usepackage[flushleft]{threeparttable}
\usepackage{caption}
\usepackage{tikz}
\newcommand{\pie}[1]{%
\begin{tikzpicture}
 \draw (0ex,0ex) circle (1ex);
 \fill (0ex,-1ex) arc (-90:(#1-90):1ex) -- (0ex,-1ex) -- cycle;
\end{tikzpicture}%
}

\def\eg{\emph{e.g.}\xspace}

\def\ie{\emph{i.e.}\xspace}

\newcommand{\one}{({\em i}\/)\xspace}
\newcommand{\two}{({\em ii}\/)\xspace}
\newcommand{\three}{({\em iii}\/)\xspace}

\DeclareMathOperator*{\argmin}{arg\,min}

\def\BibTeX{{\rm B\kern-.05em{\sc i\kern-.025em b}\kern-.08em
    T\kern-.1667em\lower.7ex\hbox{E}\kern-.125emX}}
\begin{document}

\title{PublicCheck: Public Integrity Verification for Services of Run-time Deep Models}

\author{
\IEEEauthorblockN{
Shuo Wang\IEEEauthorrefmark{1}\IEEEauthorrefmark{2},
Sharif Abuadbba\IEEEauthorrefmark{1}\IEEEauthorrefmark{2}, 
Sidharth Agarwal\IEEEauthorrefmark{3},
Kristen Moore\IEEEauthorrefmark{1}\IEEEauthorrefmark{2},
Ruoxi Sun\IEEEauthorrefmark{1}, 
\\
Minhui Xue\IEEEauthorrefmark{1}\IEEEauthorrefmark{2},
Surya Nepal\IEEEauthorrefmark{1}\IEEEauthorrefmark{2},
Seyit Camtepe\IEEEauthorrefmark{1}\IEEEauthorrefmark{2} and 
Salil Kanhere\IEEEauthorrefmark{4}}
\IEEEauthorblockA{
\IEEEauthorrefmark{1}CSIRO's Data61, Australia}
\IEEEauthorblockA{
\IEEEauthorrefmark{2}Cybersecurity CRC, Australia}
\IEEEauthorblockA{
\IEEEauthorrefmark{3}Indian Institute of Technology Delhi, India}
\IEEEauthorblockA{
\IEEEauthorrefmark{4}University of New South Wales, Australia}
}

\maketitle

\begin{abstract}
Existing integrity verification approaches for deep models are designed for private verification (\ie, assuming the service provider is honest, with white-box access to model parameters). However, private verification approaches do not allow model users to verify the model at run-time. Instead, they must trust the service provider, who may tamper with the verification results. In contrast, a public verification approach that considers the possibility of dishonest service providers can benefit a wider range of users. In this paper, we propose PublicCheck, a practical public integrity verification solution for services of run-time deep models. PublicCheck considers dishonest service providers, and overcomes public verification challenges of being lightweight, providing anti-counterfeiting protection, and having fingerprinting samples that appear smooth. To capture and fingerprint the inherent prediction behaviors of a run-time model, PublicCheck generates smoothly transformed and augmented encysted samples that are enclosed around the model's decision boundary while ensuring that the verification queries are indistinguishable from normal queries. PublicCheck is also applicable when knowledge of the target model is limited (\eg, with no knowledge of gradients or model parameters). A thorough evaluation of PublicCheck demonstrates the strong capability for model integrity breach detection (100\% detection accuracy with less than 10 black-box API queries) against various model integrity attacks and model compression attacks. PublicCheck also demonstrates the smooth appearance, feasibility, and efficiency of generating a plethora of encysted samples for fingerprinting.
\end{abstract}

\section{Introduction}\label{sec_introduction}

\begin{figure*}[t]
    \centering
    \includegraphics[width=0.98\linewidth]{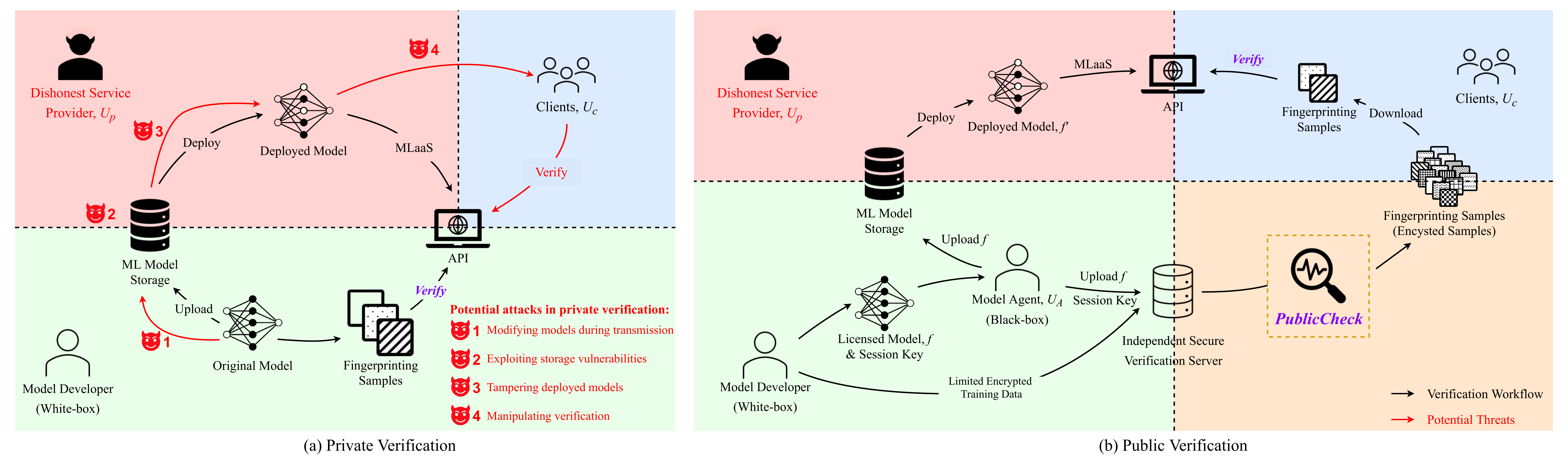}
    \caption{An overview of private verification and public verification.}
    \label{fig_overview}
\end{figure*}

Cloud-enabled Machine Learning as a Service (MLaaS)  has shown enormous promise to transform how deep learning models are developed and deployed. Model agents, who facilitate the commercialization of machine learning based frameworks or licensed products, feature in the entire pipeline - from downstream end-users (clients) to upstream cloud platforms. Their role is to commercialize pre-trained models (that they did not necessarily develop themselves) by deploying them for profit to cater to broader end users' demands. One of the benefits of such a business model is the isolation of the development, deployment, and use of models, \eg, the details of the deployed model need not be known by model agents and clients. 
For example, Atrium (atrium.ai) recently created a Machine Learning Model Broker service (model agent) to facilitate machine learning adoption. Hence, the MLaaS business ecosystem now includes four players. A model developer (the original owner), a model agent (the broker, \eg, Atrium) who facilitates model adoption without white-box access, a service provider (Google/Amazon) where the model is deployed with commercial APIs with black-box query access, and model clients who commercially use the model by querying those service provider APIs. 

On an orthogonal path, potential risks and security threats are emerging with MLaaS, since the pre-trained models can be maliciously modified through Trojan or backdoor attacks~\cite{li2021hidden,li2020invisible,zhong2020backdoor,chen2017targeted,turner2018clean,liu2017trojaning,ma2023beatrix}, or can be degraded~\cite{he2019sensitive} and even backdoored by model compression~\cite{ma2021quantization,tian2022stealthy} during deployment. To protect the model integrity and benefits of end-users, it is imperative for model agents, service providers, and end-users to verify whether the deployed model has been tampered with. 
Contrary to the traditional integrity verification process for objective entities (\eg, files hashing), this work intends to conduct \textit{integrity verification on services of the run-time deep model} in the MLaaS platform (\eg, black-box access to the prediction service). 
To achieve this, verification keys, such as fingerprinting (sample, label) pairs, are designed to verify the deployed surrogate models, derived from the source of original or licensed models (\eg, model developers or model agents). During verification, if a deployed model outputs a different label on a fingerprinting sample, the model will be considered modified. Most existing verification techniques for model integrity belong to private verifiability~\cite{adi2018turning} by honest verification providers, which only provide a verification service to honest parties, such as the model developers who have white-box access to the training data and model parameters. However, a dishonest verification service provider could easily manipulate the integrity verification results, for example, by providing a verification service through the hash value of a compromised version of the deployed model. Additionally, as the service providers themselves are part of the service, self-validation of verification results is unfair when providers can profit from being dishonest. 
Public verifiability aims to serve a wider range of users, including model agents, service providers, and clients who opted in or purchased the API. This also releases model developers from the workload of conducting or maintaining the verification service, enabling a more flexible business model through the inclusion of model agents. Most importantly, it disentangles the integrity protection of the deployed model from the trustworthiness identification of the service provider (\eg, cloud platform) (see Figure~\ref{fig_overview}).

To defeat dishonest cloud providers, the key is to guarantee that the verification process is \textbf{indistinguishable} from normal business, which forces the service provider to always answer the queries with predictions output from the actual deployed model, rather than from a compromised model. Thus, the design of the verification key must be efficiently linked to the model prediction behaviors, and should satisfy the following three requirements:

\begin{itemize}[leftmargin=*]
    \item \textit{Lightweight}  
    to ensure an affordable computational cost for a large-scale disposable (designed to be used once per user within a limited time period to avoid replay attacks) verification key generation. 
    \item \textit{Smooth appearance}  
    to ensure verification queries are indistinguishable, and to enable protection against adaptive attacks. One feasible approach is to add structural patterns in pixel space that blends in with the original manuscript. 
    \item \textit{Anti-counterfeiting} 
    to prevent the adversary from bypassing verification or training a surrogate model. Randomness and uncertainty should be incorporated into the design for secure and long-term verification.
\end{itemize}

\begin{figure}[t]
\centering  
\includegraphics[width=0.98\linewidth]{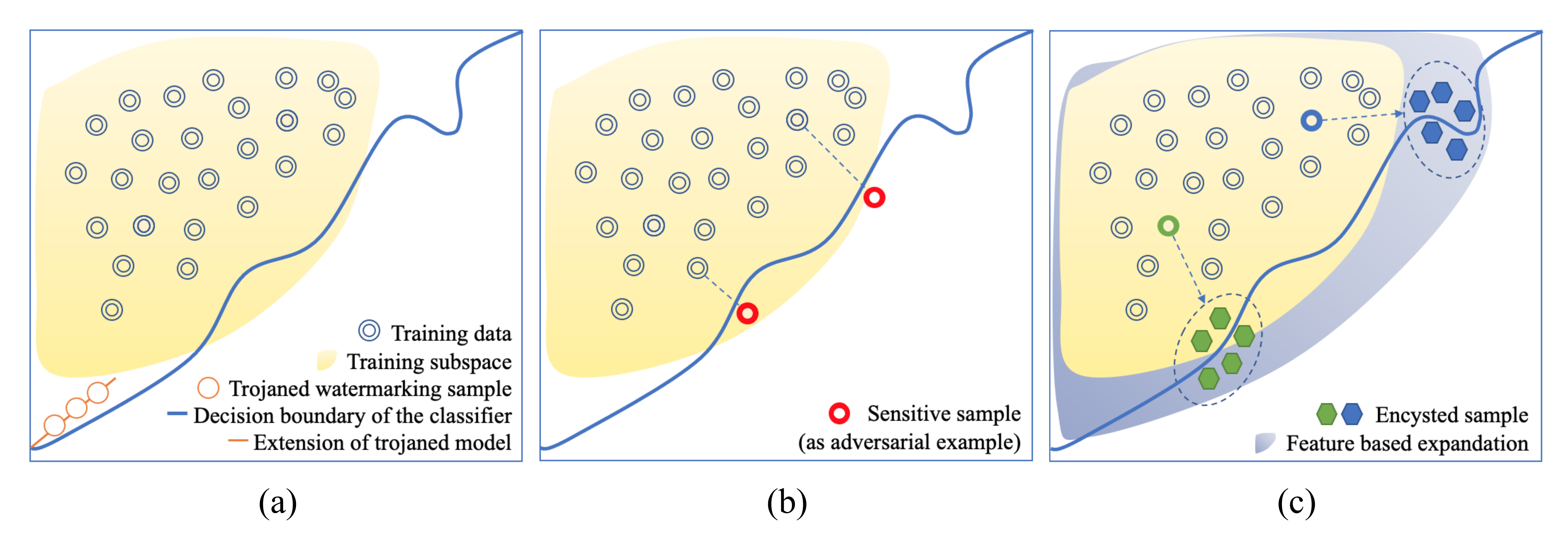} 
\caption{Demonstration of the deep neural networks fingerprinting approaches: (a) trigger-based fingerprinting; (b) adversarial-based fingerprinting; (c) PublicCheck (ours). }
\label{fig_watermarking_approaches}
\vspace{2mm}
\end{figure}

\noindent{\bf Our solution.}
The purpose of this research is to answer the following research question: \textit{how can we enable MLaaS users to publicly verify the integrity of services for the underlying deep model while satisfying the three aforementioned requirements to defeat the dishonest cloud provider?} 
To the best of our knowledge, the proposed PublicCheck is the first practical methodology for public integrity verification, with only black-box access to the target MLaaS models during both fingerprinting design and inference. The intuition is that manipulation of the model will manifest as a shift in the decision boundary, changing the prediction of certain samples in the general vicinity of the decision boundary. The key idea is to leverage a generative autoencoder model and apply clever sample selection strategies to conduct controllable data augmentation that encysts the target model's decision boundary (Figure~\ref{fig_watermarking_approaches}(c)), to capture the predicted behavior patterns around the decision boundary. A set of encysted samples and associated predictions serve as the fingerprint of the model. We also consider the design of verification samples in the scenario where there is limited knowledge (\eg, gradients or parameters) of the target model, and perform the verification with black-box API queries. This covers cases where the target models could be non-gradient-based or encapsulated as executable files, where the only available information is the prediction result. 

\begin{table*}[t]
\centering
\caption{An overview of state-of-the-art verification.}
\label{tab_watermarking_approaches}
\resizebox{0.9\linewidth}{!}{
\begin{threeparttable}
\begin{tabular}{lccccccc}
\toprule
\multicolumn{1}{l}{\multirow{3}{*}{\textbf{Verification Approaches}}} &
  \multicolumn{3}{c}{\textbf{Knowledge NOT Required}} &
  \multirow{3}{*}{\textbf{\begin{tabular}[c]{@{}c@{}}No Model \\ Degradation\end{tabular}}} &
  \multicolumn{3}{c}{\textbf{Public Verifiability Requirements}} \\ \cmidrule(lr){2-4} \cmidrule(l){6-8} 
\multicolumn{1}{c}{} &
  \begin{tabular}[c]{@{}c@{}}Gradient/\\ Parameters\end{tabular} &
  \begin{tabular}[c]{@{}c@{}}Model \\ Fine-tuning\end{tabular} &
  \begin{tabular}[c]{@{}c@{}}Backward\\ Propagation\end{tabular} &
   &
  Lightweight &
  Smooth &
  \begin{tabular}[c]{@{}c@{}}Anti-\\counterfeiting\end{tabular} \\ \midrule
\textbf{Hash-based verification}
& \pie{360} & \pie{360} & \pie{360} & \pie{360} & \pie{360}  & N/A  & \pie{0}  \\
\begin{tabular}[l]{@{}l@{}}\textbf{Backdoor-based}\\\textbf{watermarking~\cite{zhong2020backdoor,adi2018turning}}\end{tabular}
& \pie{0} & \pie{0} & \pie{0} & \pie{0} & \pie{0}  & \pie{0}  & \pie{0}  \\
\begin{tabular}[l]{@{}l@{}}\textbf{Adversarial-based}\\\textbf{watermarking~\cite{he2019sensitive,le2020adversarial,lukas2021deep} }\end{tabular}
& \pie{0} & \pie{360} & \pie{0} & \pie{36   0} & \pie{0}  & \pie{0}  & \pie{0}  \\
\textbf{PublicCheck (Ours)}
& \pie{360} & \pie{360} & \pie{360} & \pie{360} & \pie{360} & \pie{360} & \pie{360} \\ \bottomrule
\end{tabular}
\begin{tablenotes}
    \item[] \pie{0}: the property or requirement is not satisfied by the approach; \pie{360}: the property or requirement is satisfied by the approach.
\end{tablenotes}
\end{threeparttable}
}
\vspace{-3mm}
\end{table*}

\noindent{\bf Our contributions.}
The key contributions are as follows:
\begin{itemize}[leftmargin=*]
    \item We propose the \textbf{\textit{first}} public integrity verification approach for services of run-time deep models via encysted sample augmentation. The public integrity verification considers dishonest service providers. 
    
    \item We implement a lightweight generation strategy to produce large-scale verification samples in a low-cost and controllable manner in terms of semantic attributes. Randomness and uncertainty are integrated into the generation via latent perturbation, while smoothness selection is used to filter candidate samples. PublicCheck can be applied in model-agnostic scenarios using black-box knowledge of the target models, with no assumptions on network architecture, hyper-parameters, and optimization. 
    
    \item We implement and evaluate our approach against numerous integrity attacks and model compression across different models. PublicCheck demonstrates 100\% accuracy on model integrity detection, with low overhead (less than ten API queries). The appearance evaluation of encysted samples reveals that they are smooth and indistinguishable from normal ones. 
    We also demonstrate the feasibility of generating a large number of verification samples, reducing the time from 300 seconds to 1 second compared to the existing adversarial perturbation-based approaches.
\end{itemize}

\section{Related Work}
Existing verification approaches can be categorized into \textit{hash-based verification}~\cite{fridrich2000robust}, \textit{trigger-based fingerprinting}~\cite{adi2018turning}, and \textit{adversarial-based fingerprinting}~\cite{he2019sensitive,le2020adversarial,lukas2021deep}. However, none of them can fulfill the requirements of lightweight, smooth appearance, and anti-counterfeiting whilst being applied to public verification. We present an overview of state-of-the-art model verification in Table~\ref{tab_watermarking_approaches}.

\textit{Hash-based verification} is only available to entities, which can easily be tampered with. 
\textit{Trigger-based verification} produces (backdoor sample, target label) pairs to extend the classifier's decision boundary as the fingerprinting of the model (Figure~\ref{fig_watermarking_approaches}(a)). 
The generation of fingerprinting samples through model fine-tuning and backward propagation demands a significant computational resource, and in addition to this, the backdoor embedding also causes model degradation.
\textit{Adversarial-based verification} utilizes the misclassification behaviors introduced by well-designed noise in the pixel space as fingerprinting (Figure~\ref{fig_watermarking_approaches}(b)). Utilizing adversarial examples has the primary advantage of eliminating the need for training or re-training and enabling the black-box inference capability. 
Therefore, in trigger-based and adversarial-based fingerprinting approaches, white-box knowledge of the model is needed, such as gradients/parameters. Additionally, the computation cost of backward propagation is high for these approaches. 
Furthermore, these approaches only allow verification by an honest party---they are only privately verifiable due to their technical limitations of lacking randomness/uncertainty, infinite generation, or imperceptibility.
Meanwhile, the smooth appearance of fingerprinting samples (see Figure~\ref{fig_ss} in the Appendix), as well as the anti-counterfeiting protection (a malicious service provider can embed the same backdoor or conduct adversarial training to bypass verification, as demonstrated in Section~\ref{sec_evaluation_of_adaptive_attacks}), is not guaranteed. 

The trending MLaaS platforms nowadays have attracted massive users, and this necessitates the need for public verifiability with those requirements. 
To the best of our knowledge, this is the first work to promote the public integrity verification of deployed models by leveraging lightweight, smoothness, and anti-counterfeiting properties while only using black-box access during both inference and design procedures. 

\section{Problem Statement and Threat Model}
\label{sec_problem_statement}
We consider the public verification of models deployed in the cloud to provide prediction services via APIs, for any user roles, including the model agent, service providers, and any model clients. This is a typical MLaaS service provided by service providers, such as Amazon SageMaker. 
The process of the public verification is illustrated in Figure~\ref{fig_overview}(b).

\subsection{Problem Statement}
 
\noindent \textbf{Trusted third-party verification server.} The model developer will transmit the licensed model~$f$ and the encrypted sampled training data (derived from the same distribution as the training data for the model~$f$) to an independent secure verification server (\ie, a trusted third party), and passes a session key (including the decryption key) to the model agent $U_A$. The model agent $U_A$ only has black-box access to the licensed model and has no knowledge of the sampled training data on the secure verification server. The agent can request the secure verification server to decrypt the encrypted sampled training data with the decryption key, and the agent controls the launch of PublicCheck along with the session key. The once-off training of the autoencoder adopted by PublicCheck takes place on the secure verification server.

\noindent \textbf{Verification pipeline.} The verification service starts with the generation of the fingerprint of a model which is represented as $\mathbb{FM}=\{(s_i,r_i)\}_{i=1}^V,~r_i=f(s_i)$, where $\{s_i\}_{i=1}^S$ are fingerprinting samples generated for $U_A$ through PublicCheck. 
Note that $V\ll S$ is the minimum number of samples required to conduct efficient verification (around 7 in our experiments), and the pool of verification pairs, $S$, should be \textit{sufficiently large} at all times to allow for multi-time disposable public verification. An independent secure verification server would be appropriate for running and maintaining the verification production service and the pool of verification pairs to meet the lightweight, smoothness and anti-counterfeiting requirements. 
The licensed model is then uploaded to and stored on a platform maintained by the cloud service provider $U_P$, and deployed to an endpoint instance to provide an API querying service for public users. 
Considering that the deployed model $f'$ on the cloud platform may differ from the licensed model $f$ (which will be detailed in the threat model),
it is important for the model agent $U_A$, model provider $U_M$ and clients $U_C$ who purchased the API to verify whether the model $f'$ is the same as $f$. 
A set of verification keys are used to verify the integrity of the deployed model. 
During verification inference, the user obtains $\mathbb{FM}=\{(s_i,r_i)\}_{i=1}^V$ from the independent secure verification server (after the paywall), and then queries the API of the deployed model $f'$ with fingerprinting samples $\{s_i\}_{i=1}^V$. 
The model is not intact if there exists $i\in\{1,\ldots,V\}$ such that $f'(s_i) \neq r_i$.

\subsection{Threat Model}
From the integrity breach perspective, there is no assumption about how the integrity of a deployed model is compromised (\ie, no limitations are placed on the adversary, including malicious service provider).
From the integrity verification perspective, we assume that the service provider is dishonest, who may maliciously modify the model in deployment and intentionally provide counterfeit verification services.

\noindent \textbf{Adversary's goal.} 
There are three main service procedures during model deployment: \one model transition to the cloud; \two model storage in the cloud; and \three inquiry services after deployment via APIs. 
We assume that model integrity could be compromised at any procedure, and make no assumptions about how the model is modified, by an external attack or internal manipulation. 
The \textit{Integrity adversary's goal} is to derive a surrogate model to replace the original licensed model. The adversary may embed backdoor or misclassification behaviors in the modified model for malicious purposes. A dishonest cloud service provider may deploy degraded models that will incur a lower cost (\eg, model compression). The \textit{Verification adversary's goal} is to identify the verification keys from queries and then counterfeit the verification keys or the verification process to bypass a compromised model. For example, during private verification (Figure~\ref{fig_overview}(a)), only the model developers can use the service to determine whether the deployed model is intact, while they can be fooled by the counterfeit verification keys issued by the service provider. 

\noindent \textbf{Adversary's capacity.} 
We summarize both the integrity and verification adversary's capacity with respect to potential vulnerabilities and security risks in MLaaS. We also present these threats in Figure~\ref{fig_overview}(a) with red arrows. 
We note that all vulnerabilities on the path from the model fingerprinting to the model integrity verification are invisible to clients, as well as to the verification tools such as PublicCheck. Therefore, we would like to propose a single solution to all these potential vulnerabilities, rather than solve them one by one. Besides the other attackers, the dishonest cloud service provider is considered as a much stronger insider attacker. We summarize some typical examples of an adversary's capacity as follows.

\begin{itemize}[leftmargin=*]
\item \textit{Modifying models during transmission (Integrity).} The adversary can modify a model while it is transmitted from the model agent to the service provider by exploiting vulnerabilities in cloud network protocols or service interfaces, \ie, during \textit{upload} procedure in Figure~\ref{fig_overview}(a). 

\item \textit{Exploiting storage vulnerability (Integrity).} 
Adversaries can exploit \textit{machine learning model storage vulnerabilities} to substitute a compromised model for the safe one.

\item \textit{Tampering deployed models (Integrity).} The insider attacker is capable not only of backdooring the deployed model but also of other abnormal behaviors such as model compression. Note that model compression could be a weak attack, but acts as a worst-case for verification (modifications of model behavior could be hard to identify).
\item \textit{Manipulating verification (Verification).} The adversary can manipulate the verification results, either manipulating the results of verification queries or forging verification keys to convince the user that the deployed model is intact. For example, a dishonest service provider can leverage a discriminator to identify verification queries and arbitrarily manipulate the verification results before being sent to the user. Otherwise, the adversary can publish fake a hash code or embed fingerprinting samples into modified models through fine-tuning or adversarial training to bypass the verification.

\end{itemize}

\noindent \textbf{Adversary's knowledge.} 
Existing deep neural networks watermarking or fingerprinting approaches~\cite{fridrich2000robust,adi2018turning,he2019sensitive} assume the service provider is trusted. This is too strong an assumption, as the cloud provider has white-box access to all the parameters and run-time information, resulting in the potential to breach the fingerprinting patterns through the backdoor or adversarial misclassification behaviors. 
To model a more practical scenario, we assume that the adversaries, including the service provider, have white-box knowledge of the deployed models. We also assume that an adversary has white-box access to a collection of verification samples. 

\section{PublicCheck System Design}

\begin{figure}[t]
\centering  
\includegraphics[width=\linewidth]{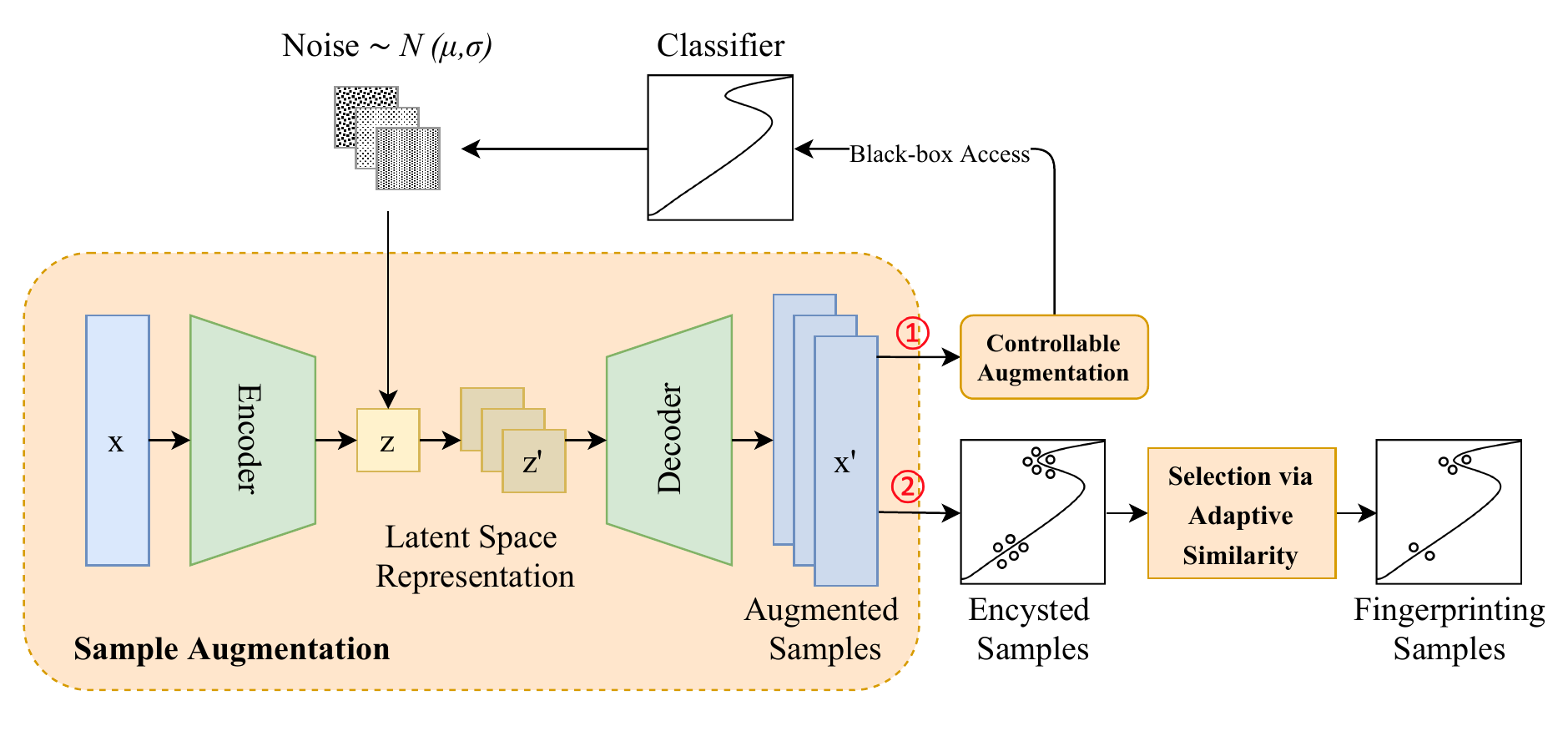}
\caption{Overview of the PublicCheck fingerprinting.}
\label{fig_overview_publicCheck}
\end{figure}

\begin{figure*}[t]
\centering  
\includegraphics[width=0.95\linewidth]{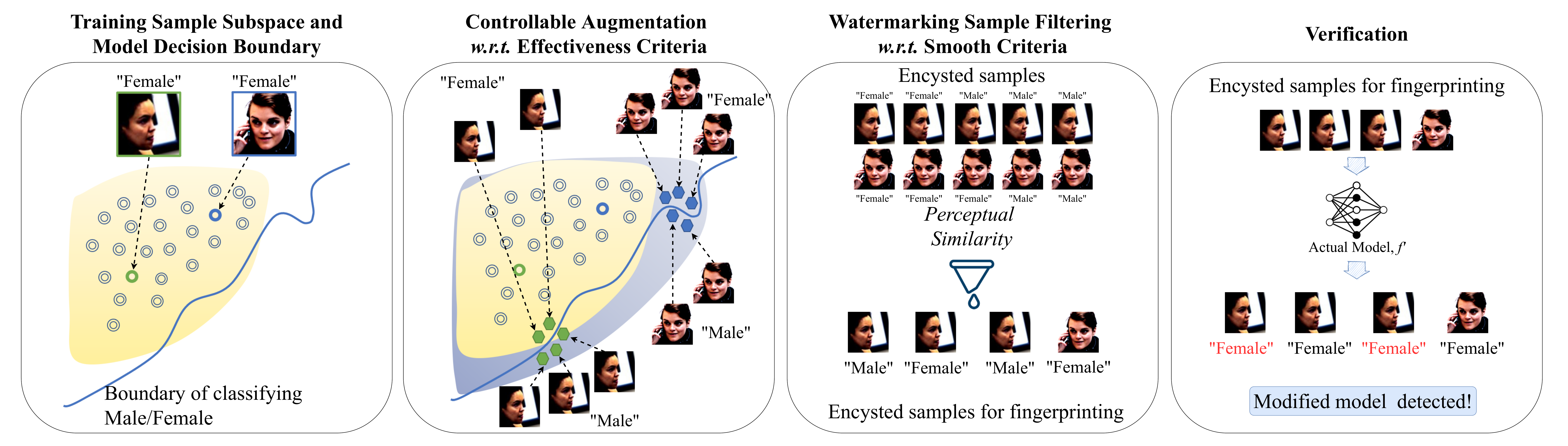}
\caption{Visual illustration of our public fingerprinting.}
\label{fig_illustrate}
\end{figure*}

PublicCheck first conducts \textit{data augmentation}. A small number of training samples are used as referenced samples ($x$), and data augmentation is achieved by reconstructing the perturbed encoded latent codes ($z$ to $z'$) after adding random noise. The noise is restricted by two criteria: \one \textit{controllable augmentation} ensures the reconstructions are densely populated around the decision boundary, and the prediction behavior of the target classifier is controlled; and \two \textit{selection of candidates based on smoothness} ensures the smooth appearance of reconstructions via adaptive similarity. After candidate selection, the reconstructed samples $x'$ are used for fingerprinting verification. 
Finally, the model's fingerprint is represented as a prediction vector for a small set of encysted samples. 
During verification inference, the model is verified to be intact when the response of the test model on the encysted sample set is equal to the fingerprint prediction vector. 
The overview of the PublicCheck verification is illustrated in Algorithm~\ref{alg_PublicCheck} and Figure~\ref{fig_overview_publicCheck}. A visual illustration is provided in Figure~\ref{fig_illustrate}. 

\subsection{Data Augmentation} 
Using attribute manipulation-based generation, we augment data samples along a semantic feature axis for efficient data augmentation. Variational autoencoder (VAE)-based generative models~\cite{kim2018disentangling,burgess2018understanding} are developed to represent samples in the high-dimensional pixel space into a low-dimensional latent space via latent codes. 
Combined with disentanglement, they are capable of providing such a controllable mechanism.  We aim to develop latent representations that encode distinct attributes of data as disentangled latent codes, where changes in one part of a latent code correspond only to changes in a single attribute of data. 
We introduce generative models with two levels of disentanglement in an unsupervised learning manner, where attribute-level disentanglement is based on a low-dimensional latent representation vector and abstraction-level disentanglement is based on high-dimensional latent representations. 

For low-fidelity and simple images, \eg handwritten digits MNIST, the VAE-based generative model can be used to learn a simple representation of data in the latent space, \eg, a 20-dimensional latent vector.  Attribute-level disentanglement can be attained by forcing the distribution of latent representations to be factorial via loss terms, leading to independent distributions across dimensions. 
To better balance the trade-offs between reconstruction quality and disentanglement, we adopt the Total Correlation (TC)~\cite{kim2018disentangling} and adopt a human perceptual evaluation Learned Perceptual Image Patch Similarity (LPIPS)~\cite{zhang2018unreasonable} as a reconstruction error (see details in Appendix~\ref{appendix_vae}). 

For high-fidelity and complicated images, such as high-resolution human face images, or multi-domain images such as CIFAR-10, the dimension of the latent codes should be largely extended to maintain more features. Therefore, the disentanglement strategies used for disentangle-VAE are not applicable. Accordingly, we provide an abstraction-level disentanglement strategy to handle high-fidelity and complicated images. We extend the scale of the latent codes and divide them into a style (high abstraction) level and a texture (low abstraction) level latent representation, via two separate encoders. 
To extend the latent representation, we apply the Vector Quantized strategy used in previous works~\cite{razavi2019generating,van2017neural}. 
In general, VQ-based autoencoders consist of three components, encoder $En$, decoder $De$, and codebook $C$. 
The codebook $C$ can be viewed as a common feature dictionary shared between the encoder and decoder, consisting of $K$ categorical embedding items with $D$ dimensions. 
The encoder is a non-linear mapping from the input instance $x$ in the pixel space to the latent representation $z_{e}(x)$, consisting of latent embedding vectors with D dimensions. Vector Quantization is to map $z_{e}(x)$ to a discrete latent matrix with each element representing the index of the nearest embedding items in the codebook for each latent embedding vector of $z_{e}(x)$. 
The decoder reconstructs back to pixel space using the queried embedding items $z_q(z)$ corresponding to the discrete latent index matrix via another non-linear function, as shown in Figure~\ref{fig_vq} (see details of VQ-based generative models in Appendix~\ref{appendix_vq}).

To achieve abstraction-level disentanglement, we apply two separate (encoder, codebook) pairs to model texture information and style information, respectively. A global encoder $En^{g}$ and codebook $C^{g}$ are applied to capture high-abstractive information, such as style, shape, and geometry, and a local encoder $En^{l}$ and codebook $C^{l}$ are used to capture low-abstractive information, such as texture, color, or background. 
Specifically, the local encoder initially maps the input instance into local latent representation $z^l_e(x)$ using $En^{l}$, followed by conducting a global encoder $En^{g}$ to map $z^l_e(x)$ into $z^g_e(x)$. Then we transfer the $z^g_e(x)$ and $z^l_e(x)$ into a discrete latent matrix $z^{g}(x)$ and $z^l(x)$ via vector quantization by nearest neighbor searching on the global codebook $C^g$ and $C^l$, respectively. $z^{l}_q(x)$ and $z^{g}_q(x)$ are the queried/retrieved embedding items according to the discrete latent matrix. The global decoder $De^g$ is then applied to recover the latent representation $z^{g}_q(x)$ back to a representation $s^l$ with the same size as $z^l_e(x)$. The composited $\hat{z}^l_q(x)=s^l+z^l_q(x)$ are fed into the decoder $De^l$. 
Finally, the local decoder $De^l$ takes as input all levels of the quantized latent representation back to the original image size. Here, we also replace the default pixel-wise reconstruction evaluation with the perceptual evaluation metric LPIPS. 

\begin{algorithm}[t]
\footnotesize
\caption{Model Integrity Public Verification} 
\label{alg_PublicCheck}
\KwIn{ES size $V$, licensed and deployed models $f$, $f'$, maximum scale and variance of perturbation $\Delta_{max}, ~\sigma$, pre-trained encoder and decoder $En$, $De$, iteration $I'$}
\KwOut{$ES$ and integrity verification result $r$} 
$X_t \gets $ Random samples from training datasets \\ 
\While{$|ES|<V$ }{
    \For{x in $X_t$}{
        $z \gets En(x)$
        \\
        \tcp { Boundary value of noise} 
        $\mu \gets argmin~\Delta z,~s.t.~\Delta z \leq \Delta z_{max}~and~f(De(\Delta z+z)) \ne f(De(z))$ 
        \\ 
        \tcp {Filtering via adaptive similarity}
        $\xi \gets$ Adaptive Perceptual Similarity Threshold  \\ 
        \While{$i \leq I'$}
        {
        $\Delta z \sim \mathcal{N}(\mu,\sigma)$\\
        \tcp{Perceptual Similarity}
        \If{$PS(De(z+\Delta z)) \geq \xi$}
        { 
            $ES_x \gets De(z+\Delta z)$\\
        }
        $i=i+1$\\
        }
        $es_x \gets$ one random sample from $ES_x$\\
        $ES\gets ES \cup es_x$ \\ 
    }
}
Fingerprinting of $\mathbb{FM} \gets f(ES)$ \\
$r\gets $ \textit{True} if $f'(ES)=f(ES)$ else \textit{False}\\
\end{algorithm}

\subsection{Controllable Augmentation}  

Controllable augmentation aims to transform a given reference sample into a set of augmented reconstructions that encyst around the decision boundary of the target model, using the pre-trained generative model with disentanglement. 
Specifically, augmentation is conducted via adding certain perturbation $\Delta z$ into specific latent codes that clearly control a semantic attribute (line 4 of Algorithm~\ref{alg_PublicCheck}), followed by reconstruction via a decoder $De(z+\Delta z)$. The perturbation is added to selected elements of a latent code vector as attribute-level disentanglement, or to the entire low abstraction latent representation as abstraction-level disentanglement.

Encysted samples are defined as reconstructed copies from perturbed latent representations, where noise is from a given distribution within a given scale (\eg, normal distribution $\mathcal{N}(\mu,\sigma)$). 
Given a target model $f(x)$, we define the marginal value of the encysted boundary (\eg, $\mu$) as the minimum value that changes the prediction (line 6). 
Scale $\sigma$ defines the encysted noise range with the maximum noise scale (upper bound $\mu+\sigma$) and the minimum scale (lower bound $\mu-\sigma$). 
Encysted samples $ES=\{es_1,\ldots,es_n\}$ are defined as reconstructed copies using the perturbation sampled from the normal distribution $\mathcal{N}(\mu,\sigma)$. Within the encysted noise range, the predictions of the reconstructed encysted samples using the perturbed latent codes are easy to be flipped. 
An encysted sample derived from $x$ that produces a changed prediction is referred to as the outer encysted sample, and vice versa, the inner encysted sample. 
Depending on the randomly sampled noise, reconstructed encysted samples are either the inner or outer samples. We develop two strategies to effectively determine the encysted noise distribution $\mathcal{N}(\mu,\sigma)$ for the generation of the encysted sample, under the limitation of the knowledge of the target model in the following parts. 
Note that the advantage of our approach is the continuous generation of fingerprinting samples up to the plethora of amounts, producing infinite encysted samples (due to cardinality of the continuum) with lightweight computational overhead via feedforward computation using a pre-trained autoencoder, and only requires black-box inference of the target model. 
Additional generation boosting strategies are given in Appendix~\ref{appendix_booster}. 
\noindent \textbf{Fingerprinting design with few training samples.}\label{black-box} 
To address the black-box knowledge setting for the target model in Section~\ref{sec_problem_statement}, we conduct the fingerprinting design with a few training samples as follows. 
The term black-box in our work is twofold: \textit{(i)} there is only black-box access to the model to perform verification; \textit{(ii)}  the design of the fingerprinting of the model is conducted with black-box knowledge about the model (see more details in Appendix~\ref{appendix_black_box_settings}). 
Given the pre-trained autoencoder, the fingerprinting design has access only to a few training samples and the prediction results of a licensed model, without access to any other knowledge about the target model, such as its parameters and structure during fingerprinting design. 
This strategy aims to determine the outer bound of noise, namely, the minimum noise scale added to the latent representation that changes the prediction of the corresponding reconstructed image.  
The perturbation of the latent representation is considered as the optimization of $\Delta Z$ using the C\&W loss~\cite{carlini2017towards}:
\begin{equation} 
\begin{aligned}
\max (0, \log~ f(De(\cdot))_{y}- \max_{c \neq y} ~\log~ f\left( De(\cdot)\right)_{c}), 
\end{aligned}
\end{equation}
where $De(\cdot)=De(Z+\Delta Z)$ and $Z$ is the latent codes for input $x$, the $f\left( \cdot \right)_{y}$ reveals that the output of the function $f$ is $y$. The minimum loss is achieved when $f(De(Z+\Delta Z) )_{y} \leqslant \max _{c \neq y}  f\left( De(Z+\Delta Z)\right)_{c}$, \ie, we reach the outer bound of the noise $\Delta Z$.
Thus, finding the outer bound of latent perturbation for the input data $x$ is defined as follows:
 \begin{equation} 
\Delta Z_{outb}=\underset{\Delta Z\leq \Delta_{max}}{\argmin} \mathcal{L}\left(De(Z+\Delta Z)\right)
\label{eq_lossz}
\end{equation}
In the black-box knowledge case, Equation~\ref{eq_lossz} is optimized via only accessing the inputs and outputs of the function $f$ (\eg, a classifier).
Therefore, the Smooth Evolution Strategies (SES) is adopted to optimize Equation~\ref{eq_lossz} via search gradients. 
We define the latent noise $\Delta Z$ to be from an isometric normal distribution with mean $\mu$ and standard deviation $\sigma$, denoted by $\mathcal{N}\left(\Delta Z \mid \mu, \sigma^{2} I\right)$, using $\iota $ as $\{\mu, \sigma\}$ \cite{li2019nattack,wierstra2014natural,mohaghegh2020advflow}. 
SES defines a search distribution $p(\cdot |\iota)$ on $\mathcal{L}\left( \cdot \right)$, $De(Z+\Delta Z)$ is denoted by $\cdot$ to simplify notation, followed by optimization under this distribution on the objective:
 \begin{equation} 
J(\iota )=\mathbb{E}_{p\left(\cdot \mid \iota\right)}\left[\mathcal{L}\left(\cdot\right)\right]=\mathbb{E}_{\mathcal{N}\left(\Delta Z \mid \mu, \sigma^{2} I\right)}\left[\mathcal{L}\left(\cdot)\right)\right]
\label{eq_lossj}
\end{equation}
Gradient descent is then used to optimize the Equation~\ref{eq_lossj} by figuring out the Jacobian of $J(\iota)$. Here, $\sigma$ is considered as a hyperparameter, and the loss $J(\cdot )$ will be only optimized with respect to $\mu$.
The parameters $\iota $ are updated via a gradient descent step with learning rate $\alpha$ in Equation~\ref{eq_black}. 
Finally, we generate the encysted samples by reconstructing the perturbed latent representation using the noise $\Delta Z$ sampled from $\mathcal{N}\left(\Delta Z \mid \mu, \sigma^{2} I\right)$. 
\begin{equation} 
\label{eq_black}
\begin{aligned}
&\iota_{\mu}  \leftarrow \iota_{\mu} -\alpha \nabla_{\mu } J(\mu,\sigma),
\\
&\nabla_{\mu } J(\mu,\sigma )=\mathbb{E}_{\mathcal{N}}\left[\mathcal{L}\left(\cdot \right) \nabla_{\iota } \log \left( \mathcal{N}\left(\Delta Z \mid \mu, \sigma^{2} I\right) \right)\right]
\end{aligned}
\end{equation}

In our case, only one parameter (\ie, perturbation $\Delta z$ for specific latent code) is needed to be learned via black-box optimization. Additionally, the sampling is guided by the latent semantic feature vector. The number of iterations is generally less than 100. A greedy algorithm can also be implemented to approximate a noise value range for latent perturbation along a specific latent semantic feature vector.

\noindent \textbf{Fingerprinting design with a substitutive model.~}
\label{gray-box}
In this scenario, the fingerprinting design is assumed to have a substitutive model with respect to the target model, even with different structures and parameters. It is practical to train such a substitutive model using similar training datasets like those used to train the target model.  
This strategy aims to obtain an approximation for the outer bound of noise using the given pair of substitutive model $\mathcal{M}$ for the target model $f$ and its attacked version $\mathcal{M^{\circ}}$. Namely, the minimum noise scale added to the latent representation distinguishes the prediction of $\mathcal{M}$ from that of $\mathcal{M^{\circ}}$ on the same reconstructed image.  
Formally, the perturbation of the latent representation is the approximation of $\Delta Z$ using loss:
\begin{equation} 
\begin{aligned}
\max (0, \log \mathcal{M}\left(De(\cdot)\right)_{y}- \max_{c \neq y} \log \mathcal{M^{\circ}}\left(De(\cdot) \right)_{c})
\end{aligned}
\end{equation}
Here, $De(\cdot)=De(Z+\Delta Z)$. The minimum loss is achieved when $\mathcal{M}(De(Z+\Delta Z) )_{y} \leqslant \max _{c \neq y} \mathcal{M^{\circ}}\left( De(Z+\Delta Z)\right)_{c}$, namely, we reach one approximation of as outer bound of the noise $\Delta Z$.
Finding the outer bound of latent perturbation for the input data $x$ is defined as:

\begin{equation} 
\Delta Z_{outb}=\underset{ \tiny {\begin{matrix}\Delta Z\leq \Delta_{max},\\
\mathcal{M}(De(Z+\Delta Z))=\mathcal{M}(De(Z))
\end{matrix}} }{\argmin } \mathcal{L}\left(De(Z+\Delta Z)\right)
\label{eq_losszg}
\end{equation}

We define the latent noise $\Delta Z$ from an isometric normal distribution with mean $\mu$ and standard deviation $\sigma$, denoted by $\mathcal{N}\left(\Delta Z \mid \mu, \sigma^{2} I\right)$. In this case, we directly use the approximation of $\Delta Z_{outb}^x$ as the $\mu^x$ for each input $x$, while $\sigma$ is considered as a hyperparameter. 
Finally, we generate the encysted samples by reconstructing the perturbed latent representation using the noise $\Delta Z$ sampled from $\mathcal{N}\left(\Delta Z \mid \mu, \sigma^{2} I\right)$, targeting function $f$. 

\subsection{Sample Selection} 
We then apply the sample selection (lines 8-14 of Algorithm~\ref{alg_PublicCheck}) to enhance the smoothness of the augmented encysted samples, using an adaptive threshold to build a pool of smooth samples for random selection. 
Following augmentation of the encysted samples using perturbation scales derived from the two strategies above, the next step is to use the filtering strategies to select suitable encysted samples to be fingerprinted to meet the smoothness criterion.
We retain only the encysted samples with a higher similarity evaluation based on LPIPS. 
A fixed threshold for perceptual similarity is difficult to establish for different data types, so we propose an adaptive approach for assessing smoothness.
Given a small set of reference instances (generally around 10) randomly selected for each class, denoted by $RI_{c}$, the LPIPS loss metric is then applied to calculate the similarity between every two instances of the same class. We use the average similarity as the adaptive threshold $\xi_c$ of each class for the filtering. 
During the filtering, given a reconstructed encysted sample candidate $es_i$, the adaptive threshold $\xi_c$ and reference instance set $RI_c$ according to its predicted class via $f(es_i)=c$ are used to decide the filtering results. Specifically, we calculate the LPIPS similarity between the $es_i$ and each sample in $RI_c$. The encysted sample $es_i$ will be selected as a fingerprinting sample only if all distances are smaller than the threshold $\xi_c$. Formally, the encysted sample for fingerprinting is satisfied only when $LPIPS(es_i, r_i) \leq \xi_c,~ \forall r_i \in RI_c$. 
Note that the sample selection can also incorporate other criteria filters as a defense against adaptive attacks, as demonstrated in Section~\ref{sec_evaluation_of_adaptive_attacks}, such as prediction confidence, discriminators, or other similarity evaluation metrics.

\section{Evaluation of PublicCheck}
 
We evaluate the performance of PublicCheck on three datasets: MNIST~\cite{lecun1998gradient}, CIFAR-10 and CIFAR-100~\cite{krizhevsky2009learning}, and FFHQ~\cite{karras2019style,karras2020analyzing}. Target classifiers are LeNet (5 layers), ResNet-18~\cite{he2016deep}, and VGG-16~\cite{simonyan2014very}, respectively, with different complexity. 
Details of the datasets and classification models used in our experiments, as well as the accuracy of the original, clean models, are in Appendix~\ref{appendix_datasets_and_settings}. 

As the representation capability is specific to modeling the domain of data, generative models are commonly evaluated from two angles: single-domain and multi-domain. 
Limited by the capability of existing generative models, our primary focus is to evaluate the performance of PublickCheck on the single domain data. Moreover, we examine PublickCheck's performance with a generative model on multi-domain data with limited quality of generation. 
Therefore, two data sources are considered for the autoencoder. 
\newline
\textit{(i)} \textbf{Public source for single domain data.} The representation capability learned from public datasets (\eg, celebrity faces) is consistent with common human and private users from the same domain. 
As the face dataset contains only celebrities, we randomly sample 20\% of training data according to identity (\eg, 20 out of 100 people) as public available data for training a generative model only, which is similar to sampling from public data. 
\newline
\textit{(ii)} \textbf{Private source for multi-domain data.} We also consider when the public dataset is unavailable or there is domain drift for multi-domain data. The generative model is trained on sampled private training data, and 20\% is used as the baseline. 
We also evaluate the performance of PublicCheck on limited categories and limited size of sampled data (5\% and 10\%) in Section~\ref{param-selection}. 
We find that the integrity verification detection performance is similar under limited data settings, while the only impacted part is reconstruction quality. 
It is still one of the frontier research topics in the deep generative model field to learn good expressive representation from cross-domain and limited data. 
Instead, we implement the quality refiner strategy to enhance the quality of reconstructed images using less sampled data, as detailed in Appendix~\ref{booster}. 

We randomly sample 20\% of training data with the same distribution to establish the training dataset for the autoencoder only for both public and private source settings, and we use the remaining 80\% data in the training and testing of the original model (classifier) that we want to protect. Therefore, the 20\% selected data are from the same distribution but unseen to the original model (classifier). 

The VAE-based generative model (attribute disentanglement) is used for MNIST, and the VQ-based one (abstraction disentanglement) is used for CIFAR and FFHQ. Architectures and hyperparameters for the generative models are given in Appendix~\ref{appendix_architectures_and_hyperparameters}. Only a small set of instances (around 20) are sampled from the remaining 80\% of the dataset, to use as the reference samples for fingerprinting design, and evaluation. 
The 80\% data split is also used to train the substitutive model in Section~\ref{gray-box}.

\subsection{Evaluation Goals and Metrics}

Our first experimental goal is to determine the accuracy of PublicCheck for integrity breach detection. We test the accuracy under various attacks, as well as under model compression. In general, the less information we can derive from the underlying target model, the more difficult it is to verify its integrity. Therefore, we assume the worst-case scenario of API access, where only the Top-1 classification label (\ie, the most likely label) can be obtained.

The model fingerprint, denoted by $(ES,Y)$, is defined to be the Top-1 classification prediction of the original model on the fingerprinting set of encysted samples $ES$. The fingerprint is created by the model agent. Given a set of fingerprinting pairs $\{(es_1,y_1), (es_2,y_2), \ldots, (es_M, y_M)\} \subset (ES,Y)$ and the predictions $\hat{Y}=\{\hat{y}_1, \hat{y}_2, \ldots, \hat{y}_M\}$ of these given encysted samples by the black-box API model, we define an integrity breach to be successfully detected if: $\exists\, \hat{y_i} \in \hat{Y}$ such that $\hat{y_i}  \neq y_i $. We report the detection rate under both a few reference samples and substitutive model knowledge cases in Section~\ref{sec_evaluation_on_integrity_breach_detection}. 
We investigate the performance of our PublicCheck under more restricted scenarios in Section~\ref{param-selection}, in which the availability of data is limited.
The overheads associated with designing and implementing fingerprinting samples are in Section~\ref{sec_noise}. 
We evaluate the performance of our method when varying the hyperparameters, such as noise scale, in Section~\ref{run-time}. 
The smoothness of the encysted sample for fingerprinting is evaluated in Section~\ref{sec_evaluation_of_smoothness}. Finally, we evaluate the performance of our fingerprinting approach under an adaptive attack in Section~\ref{sec_evaluation_of_adaptive_attacks}. 
We compare our approach to the baseline (randomly selected original reference samples) and the adversarial-based approaches (\textit{Sensitive Sample}~\cite{he2019sensitive} as a typical work) for the private verification of model integrity, with white-box knowledge on target models.
\subsection{Experimental Setups}
We mimic two adversarial settings of the integrity breach in the experiments. The first is \textit{backdoor attacks} that compromised the model, assuming the target model is known and under the attacker's control. The second scenario we investigate is model compression as the hard case for verification (\ie, least changes in models). 
\textit{(i)} For the \textit{backdoor attacks}, we use the TrojanZoo~\cite{pang2020trojanzoo} platform to deploy the BadNet~\cite{gu2017badnets}, TrojanNN~\cite{liu2017trojaning}, and CleanLabel~\cite{turner2018clean} attacks on the image classification models, which are typical backdoor attacks that compromise the training data to change the model’s behavior at test time. 
In our implementation, the attack success rate of each of these 3 adversarial models was more than 95\%. The default settings of these attacks are given in Table~\ref{tab_attack_settings} in the Appendix.  
\textit{(ii)} We apply weight pruning to compress the models on three datasets. 
Pruning starts by learning the connectivity via regular network training. Next, all connections with weights below a threshold are removed, followed by retraining the network to learn the final weights for the remaining sparse connections. 
The network is pruned by retaining only important connections, with between 15\% of weights removed for the LeNet model on the MNIST dataset, 20\% removed for the ResNet model on the CIFAR-10 dataset, and around 24\% removed for the VGG16 model on the FFHQ dataset. Pruning resulted in only a minor change in the accuracy of each model ($< 2\%$). 

\begin{figure}[t]
\centering  
\includegraphics[width=\linewidth]{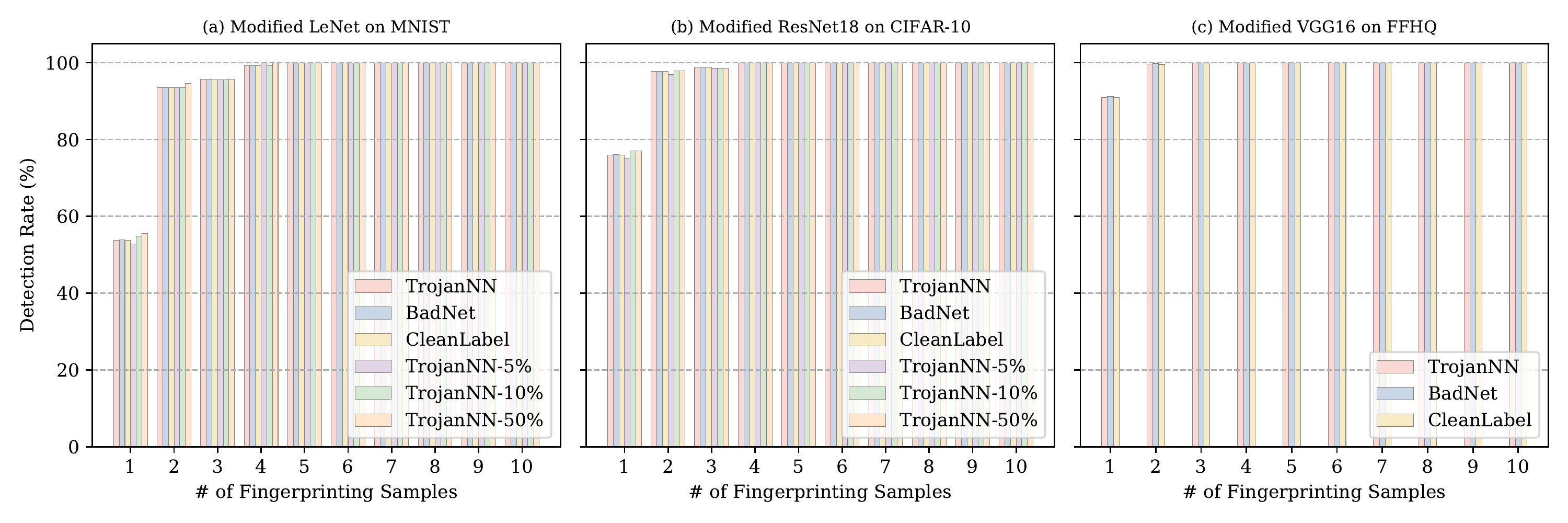} 
\caption{Detection rates for model integrity breaches under three attacks (TrojanNN, BadNet, and Clean-Label) in the few sample knowledge setting. }
\label{fig_dr_three_attacks}
\end{figure}

\subsection{Evaluation on Integrity Breach Detection}
\label{sec_evaluation_on_integrity_breach_detection}

\subsubsection{Evaluating the black-box setting of fingerprinting design with few training samples}
Under this setting, the design of the encysted sample for fingerprinting is only based on the encysted noise that is restricted by $\mu$ and user-specific scale $\sigma$ described in Section~\ref{black-box}. In our experiment, we set the default perturbation scale as $\sigma = 0.05$. 

Results of the detection success rate against three attacks are reported in Figure~\ref{fig_dr_three_attacks}. 
As shown, the detection rate rises when increasing the number of encysted samples used for verification. 
Even when only 2 encysted samples were used for verification, the detection rate of integrity breaches was above $93\%$ for MNIST, $97.9\%$ for CIFAR-10, and almost $100.0\%$ for FFHQ. 
Our results also suggest that the detection rate increases when the target model grows in size and complexity. 
The reason is that the complex model has a more complicated decision boundary, and our method has more sources to design the fingerprinting. 
With \textit{5 encysted samples for fingerprinting, we achieve 100\% detection accuracy for all the datasets and attacks we tested}. Furthermore, the overhead of model integrity validation is minimal, with only 5 API queries.

\begin{figure}[t]
\centering  
\includegraphics[width=\linewidth]{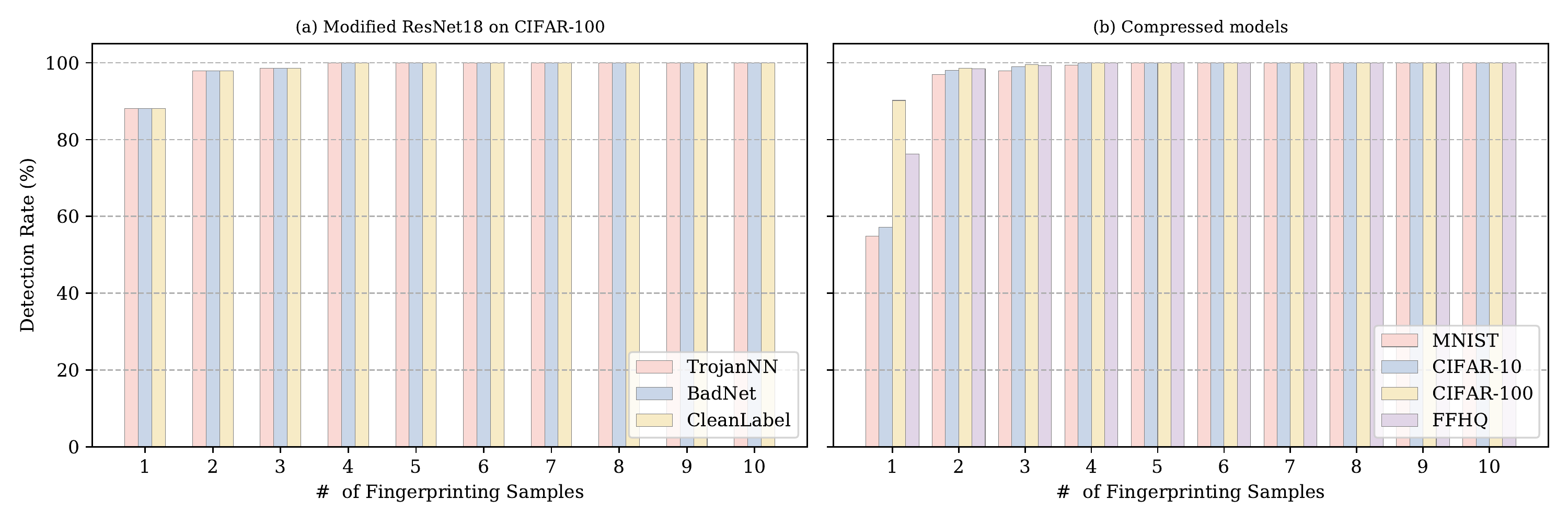}
\caption{Success detection rates for model integrity breaches (a) under three attacks for CIFAR100 and (b) under model compression on MNIST, CIFAR-10, CIFAR-100, and FFHQ datasets under rare sample knowledge settings. The axes are the number of encysted samples for fingerprinting and the detection rate.}
\label{fig_dr_larger_n_classes}
\end{figure}

To evaluate the effectiveness of our fingerprinting approach on a dataset with a larger number of classes, we also applied an evaluation on CIFAR-100. As shown in Figure~\ref{fig_dr_larger_n_classes}(a), using only 1 encysted sample for fingerprinting, the detection rate of an integrity breach was above $88\%$ for CIFAR-100 for all three attacks, as opposed to $76\%$ for CIFAR-10. Using 2 encysted samples increased the detection accuracy for CIFAR-100 to $98\%$, and using 4 samples resulted in $100\%$ accuracy. 
As a result of the more complex decision hyperplane, the detection rate increases when the classification task has more classes.

Besides the above three adversarial attacks, we also evaluate the breach detection accuracy under model compression, as shown in Figure~\ref{fig_dr_larger_n_classes}(b). 
Using only 2 encysted samples for fingerprinting, the integrity breach detection rate for the compressed models was around $97\%$ for MNIST, $98.0\%$ for CIFAR-10 and FFHQ, and 98.6\% for CIFAR-100. 
We again observe that the detection rate increases with model complexity. 
Using 5 encysted samples to fingerprint the model, we achieved 100\% detection accuracy for all datasets we tested against model compression, demonstrating the minimal overhead associated with verification. 
For the dataset with a larger number of classes, \eg, CIFAR-100, using only 1 encysted sample for fingerprinting, the detection rate of the integrity breach could reach more than $90\%$ under model compression, compared to $57\%$ for CIFAR-10. 4 encysted samples of CIFAR-100 are able to achieve 100\% detection accuracy. This also confirms the effectiveness of our fingerprinting approach when applied to a dataset with more classes. 

\noindent \textbf{Comparison with baseline.} 
When using \textit{randomly selected samples} instead of our approach to verifying three modified models attacked by TrojanNN, BadNet, and CleanLabel, we found the detection rates are around 10-20\%, which are even lower on compressed models (between 10-15\%).

\noindent \textbf{Comparison with SOTA.} In this section we compare our approach to the \textit{Sensitive Sample}~\cite{he2019sensitive} as the state of the art (SOTA) for the private verification of model integrity. 
For the detection of adversarial attacks, the performance of our PublicCheck under the few sample knowledge setting is similar to that of the Sensitive Sample under the white-box knowledge setting. For model compression, the change of parameter values under model compression is intuitively more significant than the change under attacks, since a large fraction of the parameters is removed, with values reduced to zero instead of being just slightly modified. Therefore, one would anticipate that the detection rate for model compression is higher than for adversarial attacks, which is demonstrated by our results above. 
 
We next compare the performance of PublicCheck and sensitive samples~\cite{he2019sensitive} under the same model compression settings on the CIFAR-10 classifier. Using only two samples, PublicCheck achieves a $98\%$ detection rate for CIFAR-10, compared to around $84.4\%$ for sensitive samples, and below 10\% for the randomly selected training samples. 5 samples are sufficient to achieve 100\% detection for our approach. In contrast, 10 samples are not enough to achieve 100\% detection rate for the sensitive sample approach.

It is interesting to note that the sensitive samples approach performs much worse on model compression than on attacks. This has demonstrated the major limitation of the sensitive samples approach under the public verification of the model integrity scenario. The reason is due to the design of the adversarial perturbation used to reveal the changes in the parameters of the target model. Namely, because it is based on the activation or gradient of parameter updates during training. At the same time, the model compression approaches such as pruning also utilize similar signals to remove parameters with lower activation, resulting in degradation of detection of the model change. Due to this, the gradient-based perturbation for fingerprinting was not able to detect the model changes caused by attacks or compression approaches that utilize similar gradient information. 

These results indicate that our PublicCheck can achieve state-of-the-art performance for model integrity verification against diverse attacks and model compression under rare sample knowledge settings. The advantage of our approach is that a high detection accuracy can be achieved without requiring any knowledge of model parameters or structure and with a lower cost associated with designing fingerprint samples---only feedforward procedures are required.

\subsubsection{Evaluating the black-box setting of fingerprinting design with substitutive model training}

\begin{figure}[t]
\centering  
\includegraphics[width=\linewidth]{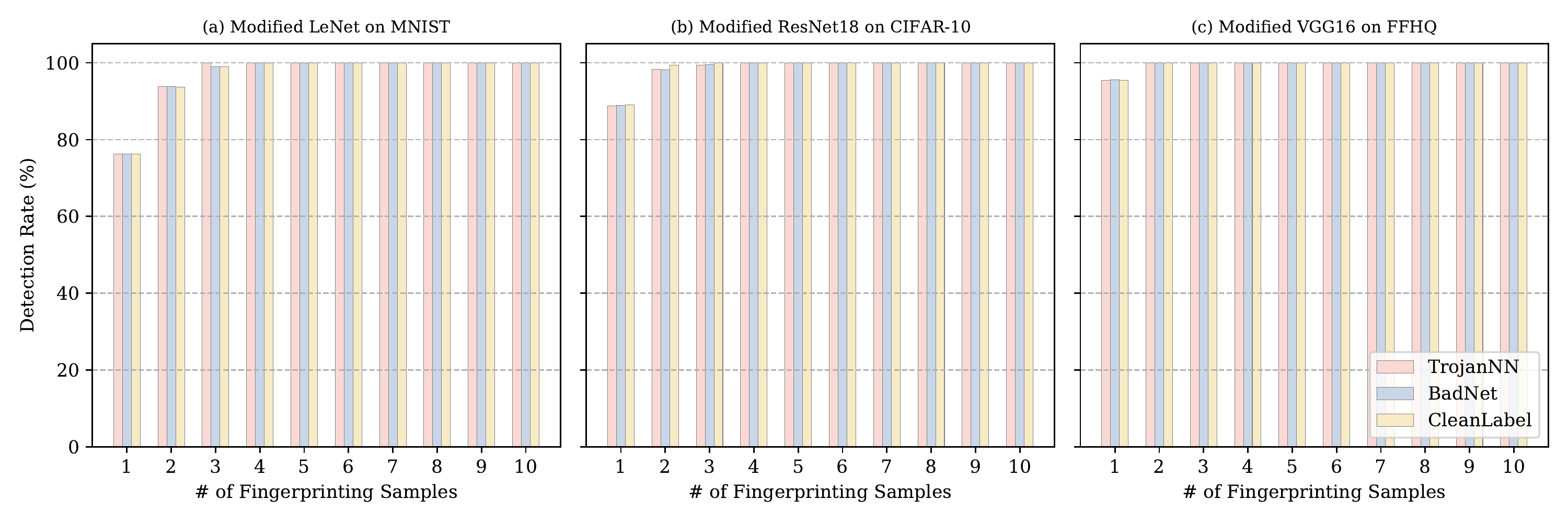}
\caption{Success detection rate for model integrity breaches under three attacks (TrojanNN, BadNet, and Clean-Label) under substitutive model knowledge setting.  The axes are the number of encysted samples for fingerprinting and the detection rate (\%).}
\label{fig_dr_substitutive}
\end{figure}

In this section, we evaluate PublickCheck when more information about the model is available. 
Under this setting, the design of the encysted sample for fingerprinting is based on the substitutive model of the target model and its attacked versions (Section~\ref{gray-box}). 
Figure~\ref{fig_dr_substitutive} shows that, as in the few sample setting, the success detection rate rises when increasing the number of encysted samples for fingerprinting. 
When using only one encysted sample for fingerprinting, the breach detection rate against adversarial attacks is higher in the substitutive model setting than in the few sample setting, by $22\%$  for MNIST, and roughly $11\%$ for CIFAR-10. 
When using two or more encysted samples, the substitutive model setting continues to outperform the few sample settings by around 1\% until both approaches achieve 100\% detection accuracy when using 5 encysted samples for fingerprinting. This was observed for all datasets and attacks tested. 
Additionally, we also evaluate the successful detection rate of model integrity breaches caused by model compression under the substitutive model setting. 
Using only one encysted sample for fingerprinting, the detection rate of the integrity breach increased by 15\% to 20\% in the substitutive model setting compared to the few sample knowledge setting.

Our results indicate that our PublicCheck can perform better against diverse attacks and model compression under substitutive model settings compared to the few sample knowledge settings, randomly selected training samples, and sensitive samples, particularly in the model compression case. 
The results demonstrate that the detection rate increases with more information about the target model, when substitute models are available. 
As 5 encysted samples for fingerprinting can obtain the 100\% accuracy of model integrity verification for both the few samples and the substitutive model knowledge settings, the former seems to be the more reasonable choice due to no assumption on the model itself. These results also demonstrate the advantage of our PublicCheck in terms of \textit{practicality} and performance under \textit{under limited knowledge}.

\subsection{Performance under Restricted Setups}\label{param-selection}

We examine performance under more restrictive conditions of the training dataset for the generative model, \ie limited categories and limited size of training samples.

\noindent \textbf{Limited categories of available samples.} 
The default assumption for the few sample knowledge settings (used in the previous section) is that the set of reference samples from the training data that are used for creating the encysted samples contain most of the classes in the dataset. However, in practice, it may not always be possible to train generative models with samples from all classes. Here we examine this more limited setting in which a small set of reference samples covering only a few categories is used. 
In the following, we report our investigation on the impact of the limited categories on integrity violation detection accuracy. 
We compare the performance of PublicCheck in detecting attacked models on MNIST and CIFAR-10 when encysted samples are generated by using only a subset of classes, namely, by randomly sampling 1, 2, 3, or 5 of the classes, instead of using all ten classes in each dataset.

\begin{table*}[t]
\centering
\caption{{Performance of PublicCheck under restricted setups. The detection rates of PublicCheck under 1, 2, 3, or 5 classes restriction have a slight drop, compared to the all-classes available scenario, demonstrating that even when watermark samples are created for 1 class only, PublicCheck still achieves 100\% accuracy when using 7 watermark samples for verification.
}}
\label{tab_class_accuracy}
\resizebox{\linewidth}{!}{
\begin{tabular}{cllllllllll}
\toprule
\multirow{2}{*}{\textbf{\# of Classes}} & \multicolumn{10}{c}{\textbf{Detection Rate ($mean \pm std$) \textit{w.r.t.} \# of Encysted Samples}} \\ \cmidrule{2-11}
  &
  \multicolumn{1}{c}{\textbf{1}} &
  \multicolumn{1}{c}{\textbf{2}} &
  \multicolumn{1}{c}{\textbf{3}} &
  \multicolumn{1}{c}{\textbf{4}} &
  \multicolumn{1}{c}{\textbf{5}} &
  \multicolumn{1}{c}{\textbf{6}} &
  \multicolumn{1}{c}{\textbf{7}} &
  \multicolumn{1}{c}{\textbf{8}} &
  \multicolumn{1}{c}{\textbf{9}} &
  \multicolumn{1}{c}{\textbf{10}} \\
  \midrule
1 & $59.0 \pm 8.2\%$ & $ 92.1 \pm 2.3\%$ & $ 94.1 \pm 2.1\%$ & $ 97.8 \pm 0.4\%$ & $ 99.2 \pm 0.3\%$ & $ 99.8 \pm 0.1\%$ & $ 100 \pm 0.0\%$ & $ 100 \pm 0.0\%$ & $ 100 \pm 0.0\%$ & $ 100 \pm 0.0\%$ \\
2 & $59.0 \pm 8.2\%$ & $ 94.4 \pm 2.0\%$ & $ 94.8 \pm 1.9\%$ & $ 98.1 \pm 0.4\%$ & $ 99.8 \pm 0.1\%$ & $ 100  \pm 0.0\%$ & $ 100 \pm 0.0\%$ & $ 100 \pm 0.0\%$ & $ 100 \pm 0.0\%$ & $ 100 \pm 0.0\%$ \\
3 & $59.0 \pm 8.2\%$ & $ 94.6 \pm 1.9\%$ & $ 96.7 \pm 0.5\%$ & $ 99.1 \pm 0.3\%$ & $ 99.9 \pm 0.1\%$ & $ 100  \pm 0.0\%$ & $ 100 \pm 0.0\%$ & $ 100 \pm 0.0\%$ & $ 100 \pm 0.0\%$ & $ 100 \pm 0.0\%$  \\
5 & $59.0 \pm 8.2\%$ & $ 94.8 \pm 1.9\%$ & $ 96.8 \pm 0.5\%$ & $ 99.3 \pm 0.2\%$ & $ 100 \pm 0.0\%$ & $ 100 \pm 0.0\%$ & $ 100 \pm 0.0\%$ & $ 100 \pm 0.0\%$ & $ 100 \pm 0.0\%$ & $ 100\pm 0.0\%$  \\ \bottomrule
\end{tabular}
}
\vspace{-5mm}
\end{table*}

Next, we evaluate whether the model integrity breach detection rate could be 100\% within the limited number of API queries. Results are averaged over these two datasets and all 3 TrojanZoo attacks. We report the results in Table~\ref{tab_class_accuracy}. 
In general, we observe an increase in accuracy as the number of classes increases. 
Although the detection rate of our PublicCheck under 1, 2, 3, or 5 classes restriction has a slight drop compared to all-classes available scenario, it is demonstrated that even when fingerprinting samples are created for one class only, PublicCheck still achieves 100\% accuracy when using seven fingerprinting samples for verification. 
It shows that 5 samples are enough to achieve 100\% detection accuracy using PublicCheck, assuming the reference samples used to generate encysted samples for fingerprinting consist of five randomly selected classes. Furthermore, the variance in detection accuracy decreases as the number of encysted samples increases. Increasing the number of different classes used in the encysted samples had little impact on the variance. 
It demonstrates our approach's generalization ability,  relying on the well-encysted and fine-granularity measure of the decision boundary using the attributed augmentation in the latent space. 
Specifically, due to the structural perturbation of PublicCheck, each encysted sample is generated from a randomly sampled perturbation, which lies very close to a random part of the decision boundary of the classification model. The prediction on the encysted sample for fingerprinting is used to distinguish whether the boundary of the verified model has the right disturbed in that localized region. Every additional fingerprinting sample then may test a different localized region of the decision boundary, thereby increasing the detection ability of PublicCheck.

\noindent \textbf{Varying sampled data for training the autoencoder.} We further evaluate the performance of PublicCheck under restricted training data for the autoencoder. In particular, we evaluate the performance of PublicCheck on less sampled data (5\%, 10\%) compared to the 20\% baseline. 
As shown in Figure~\ref{fig_dr_three_attacks}, the detection accuracy results on 5\% and 10\% size of sampled data are similar to the 20\% baseline (achieve 100\% accuracy with sample verification for both 5\% and 10\% less sampled data settings, and only within 2\% drop with less than $5$ samples verification). 
We find that the impacted part is the quality of the reconstructed images. The reconstruction error (MSE) of the 10\% and 5\% size setting increases by 39.2\% and 63.3\% respectively, compared to the 20\% baseline, as shown in Figure~\ref{appendix_mse} in the Appendix. 
However, after applying our quality booster strategy, the reconstruction error from the 5\% and 10\% size settings is equivalent to the 20\% baseline and 50\% size settings, respectively. 
After applying the quality refiner, the reconstruction distribution of the 10\% and 5\% size settings are similar to 20\% baselines, as shown in Figure~\ref{appendix_quality} in the Appendix.

\noindent \textbf{Varying the number of reference samples.} In this work, 20 randomly sampled references can generally achieve 100\% detection accuracy with 5 verification samples. The 5-sample accuracy values for [2, 4, 6, 10] reference samples are from 99\% to 100\%, while the 7-sample accuracy is all 100\%. The number of reference samples is mainly used in the smoothness filtering, which may be further reduced as the generative ability increases.

\begin{table*}[t]
\centering
\caption{{
Performance of PublicCheck when varying noise scale. Even under the broader noise scale (\ie, 0.10 or 0.50), 7 or 9 encysted samples for fingerprinting are sufficient to achieve 100\% detection accuracy.  
}}
\label{tab_noise_accuracy}
\resizebox{\linewidth}{!}{
\begin{tabular}{cllllllllll}
\toprule
\multirow{2}{*}{\textbf{Noise Scale $\sigma$}} & \multicolumn{10}{c}{\textbf{Detection Rate ($mean \pm std$) \textit{w.r.t.} \# of Encysted Samples}} \\ \cmidrule{2-11}
  &
  \multicolumn{1}{c}{\textbf{1}} &
  \multicolumn{1}{c}{\textbf{2}} &
  \multicolumn{1}{c}{\textbf{3}} &
  \multicolumn{1}{c}{\textbf{4}} &
  \multicolumn{1}{c}{\textbf{5}} &
  \multicolumn{1}{c}{\textbf{6}} &
  \multicolumn{1}{c}{\textbf{7}} &
  \multicolumn{1}{c}{\textbf{8}} &
  \multicolumn{1}{c}{\textbf{9}} &
  \multicolumn{1}{c}{\textbf{10}} \\
  \midrule
0.01 & $65.1 \pm 7.2\%$ & $95.7\pm 1.2\%$ & $97.9\pm 0.4\%$ & $99.9\pm 0.1\%$ & $100\pm 0.0\%$  & $100\pm 0.0\%$  & $100\pm 0.0\%$  & $100\pm 0.0\%$  & $100\pm 0.0\%$ & $100\pm 0.0\%$ \\
0.05 & $59.0\pm 8.0\%$ & $94.9\pm 1.8\%$ & $96.8\pm 0.5\%$ & $99.4\pm 0.2\%$ & $100 \pm 0.0\%$ & $100 \pm 0.0\%$ & $100 \pm 0.0\%$ & $100 \pm 0.0\%$ & $100\pm 0.0\%$ & $100\pm 0.0\%$ \\
0.10  & $44.5 \pm 9.2\%$ & $87.9\pm 3.1\%$ & $90.4\pm 2.5\%$ & $93.6\pm 2.1\%$ & $95.5\pm 1.2\%$ & $99.5\pm 0.2\%$ & $100\pm 0.0\%$  & $100\pm 0.0\%$  & $100 \pm 0.0\%$& $100 \pm 0.0\%$\\
0.50  & $30.3\pm 9.9\%$ & $53.4\pm 8.2\%$ & $66.4\pm 7.5\%$ & $79.8\pm 5.4\%$ & $84.3 \pm 3.9$& $92.4\pm 2.3\%$ & $95.7\pm 1.1\%$ & $98.7\pm 0.3\%$ & $100\pm 0.0\%$ & $100 \pm 0.0\%$\\ \bottomrule
\end{tabular}
}
\vspace{-5mm}
\end{table*}

\subsection{Performance across Various Noise Scales} \label{sec_noise}

To conduct the augmentation, we add a perturbation to the latent code that controls specific attributes. 
Generally, the generation of encysted samples aims to produce inner and outer encysted samples as close to the decision boundary as possible, via reconstructing the selected latent code after adding randomly sampled noise from the given distribution $\mathcal{N}(\mu,\sigma)$ under the few sample knowledge settings. 
As the noise scale $\sigma$ should be decided in advance as a hyperparameter, we now investigate \textit{``How does scaling the noise impacts the accuracy?''}. 
Intuitively, a smaller noise scale, $\sigma$, could result in a better performance of the fingerprinting samples, since more minor noise leads to inner and outer samples being closer to the decision boundary. 
We conduct experiments to evaluate the detection rate of our method when varying the noise scale $\sigma$. The results are reported in Table~\ref{tab_noise_accuracy}, confirming the intuition above that a smaller noise scale brings a higher detection rate. 
Even under the broader noise scale, \ie, 0.1 or 0.5, six or nine encysted samples for fingerprinting are sufficient to achieve 100\% detection accuracy, demonstrating the effectiveness of our method with the flexible settings. 
Furthermore, the variances of detection accuracy against different attacks decrease as the number of encysted samples increases. 
The variance of detection accuracy is also reduced as the scale of noise decreases. 
Generally, there is a trade-off between accuracy and overheads of computation. If smaller steps are taken, it takes more iterations to find the marginal value of the encysted boundary (\eg, $\mu$). Accordingly, a broader noise scale could improve the performance of a fast and flexible fingerprinting design, which is suitable for resource-constrained devices. We demonstrate this argument in the next section with respect to the run-time overheads.

\subsection{Overheads for Fingerprints Design}\label{run-time}

We examine the run-time as an overhead measure to demonstrate the practicality of PublicCheck. We mimic the end-users as a laptop with GPU resources (NVIDIA Quadro RTX 4000 8GB and i7 9900 16G CPU), and the other has CPU resources only (i5 8265U 8GB CPU). 
Two types of consumed time are examined. One is the generation of one individual encysted sample, and the other is the generation and selection of one smooth encysted sample (which we refer to as the generation of one smooth encysted sample). 
The run-time to generate a single encysted sample for fingerprinting for three given datasets under various noise scale settings is reported in Table \ref{tab_noise_time}. Given the once-off trained generator, the encysted noise range, and the default noise level of 0.05 used in our experiments, the generation time of each encysted augmentation is less than 2s for all testing datasets, including the high-fidelity images such as FFHQ images.
For MNIST and CIFAR datasets, the time values are 0.33s and 0.83s for each encysted augmentation, respectively. 
We also report the generation time for each augmentation of high-fidelity FFHQ images for CPU-Only end users, which is around 15s per augmentation. 

To ensure the smooth appearance requirement under public verification, a further filtering procedure is conducted on the set of augmented encysted samples. The overhead for the filtering procedure is minimal, less than 0.2s for selecting one smooth encysted sample for fingerprinting. The total time for the combination of augmentation and filtering of one encysted sample for fingerprinting varies from roughly 2s for MNIST, CIFAR-10, and CIFAR-100, to roughly 9s for FFHQ, under 0.05 noise level for GPU end users. For CPU-only users, run-time is similar to GPU users' for simple cases like MNIST and CIFAR-10 but can be longer for tasks involving larger class numbers or high-fidelity inputs. A possible solution for CPU users is to perform parallel processing for LPIPS evaluation across classes and the generation of encysted samples. 
Once fingerprinting samples have been generated, the time required to verify a deployed model is less than 0.2s per sample. Therefore, we can expect model verification to take less than one second when we have five fingerprinting samples. Thus, our solution can be deployed in real-time applications.

There is a trade-off between noise scale and overheads of computation. 
The broader the noise scales, the less time is used to conduct the encysted samples for fingerprinting. To achieve more efficient verification, smaller noise scales must be used, which means more iterations are needed to determine the marginal value of the encysted boundary. As demonstrated in Section~\ref{sec_noise}, even under the larger noise scale, \ie, 0.5, nine encysted samples for fingerprinting are sufficient to achieve 100\% detection accuracy, demonstrating the effectiveness of our method with the flexible settings. In this setting, the generation of the encysted samples for fingerprinting with smoothness filtering takes around 1 second for all three datasets, including high-fidelity images.

\noindent \textbf{Comparison with SOTA.} Compared to \textit{Sensitive Sample}~\cite{he2019sensitive}, we tested the time to generate sensitive samples on the same dataset, taking human face images as an example under the GPU settings. It took 2965s to generate 100 samples, and then an additional 725s to select the best 10 examples from those 100 according to its Maximum Active-Neuron Cover Sample Selection, giving a total time of 3690s to generate 10 samples. Using a filtering ratio of selecting the 10 best samples from a pool of 100 generated samples, the time to generate a single fingerprinting sample is roughly 369s, compared to around 1s for our PublicCheck. 
Reducing this filtering procedure would reduce the generation time at the cost of degradation of the performance for the selected sensitive samples. However, the time to generate one candidate of the sensitive sample (with no filtering) is still of the order of 30s, revealing that the generation of PublicCheck is still much faster. 

\begin{table}[t]
\centering
\caption{{Time(s) required to generate a smooth watermarking sample for various noise levels using GPU/CPU.}}
\label{tab_noise_time}
\resizebox{0.9\linewidth}{!}{
\begin{tabular}{lcccc}
\toprule
\multirow{2}{*}{\textbf{Datasets}} & \multicolumn{4}{c}{\textbf{Execution time under GPU/CPU settings (second)}} \\ \cmidrule{2-5}
  &
  \multicolumn{1}{c}{\textbf{$\sigma = 0.50$}} &
  \multicolumn{1}{c}{\textbf{$\sigma = 0.10$}} &
  \multicolumn{1}{c}{\textbf{$\sigma = 0.05$}} &
  \multicolumn{1}{c}{\textbf{$\sigma = 0.01$}} \\
\midrule
\textbf{MNIST} & 0.5 / 1.3 & 1.2 / 2.0  & 2.3 / 2.3  & 7.1 / 20.1  \\
\textbf{CIFAR-10} & 0.6 / 0.7 & 1.3 / 1.5 & 2.1 / 2.9 & 7.8 / 13.3  \\
\textbf{FFHQ}  & 1.2 / 15.0 & 4.8 / 25.0  & 9.1 / 59.0  & 44.4 / 230.0\\
  \bottomrule
\end{tabular}
}
\end{table}

\subsection{Evaluation of Smoothness}
\label{sec_evaluation_of_smoothness}
\begin{figure}[t]
\centering  
\includegraphics[width=0.9\linewidth]{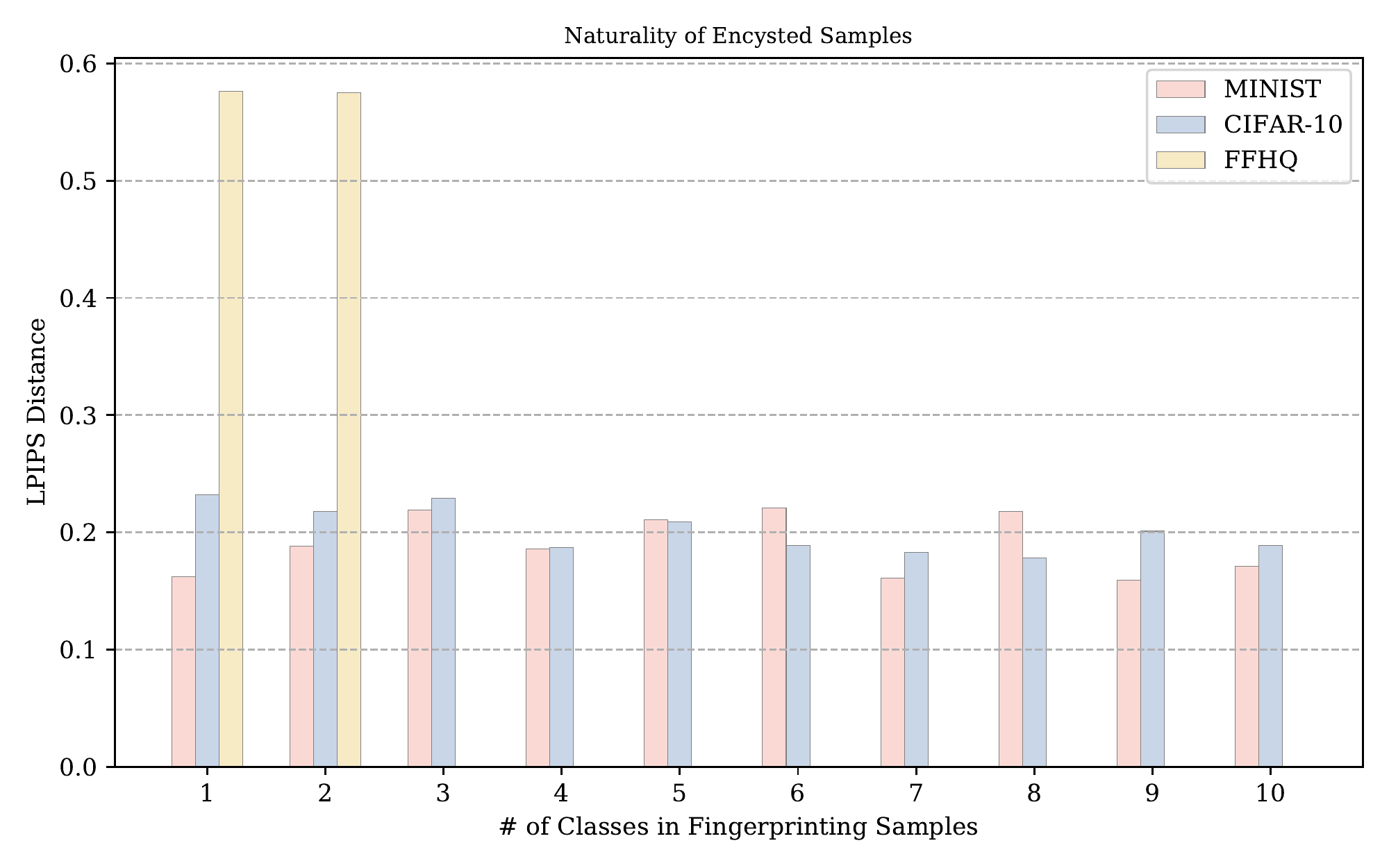}
\caption{Smoothness of encysted samples for fingerprinting among different classes.}
\label{fig_smoothness}
\end{figure}

We demonstrate the smoothness of the generated encysted sample for fingerprinting in this section. 
We first visualize the produced encysted samples for fingerprinting in Figure~\ref{fig_encysted_samples} in the Appendix. 
As shown, there are few artifacts introduced in the encysted samples compared to the original reference samples. Generally, the generated samples' small texture or color temperature is modified by manipulating the corresponding latent codes. Thus, from human perception,  these generated encysted samples are smooth. 
These results confirm the advantages of our method under the public verification scenarios. 
In addition to this, we also report the quantitative evaluation of the smoothness of encysted samples for fingerprinting, using the LPIPS metric in Figure~\ref{fig_smoothness}. 
The average LPIPS values of the encysted samples for MNIST and CIFAR-10 among classes are approximately 0.2, 0.45, and 0.55 for CIFAR-100 and FFHQ, respectively. The smoothness of encysted samples for fingerprinting is obvious and consistent among different classes for all these three datasets (FFHQ is a male and female class), demonstrating that they are smooth in pixel space. We provide more smoothness evaluation toward indistinguishability between encysted and normal samples in the following part. 

\noindent \textbf{Comparison with SOTA.} Roughly 25\% of randomly augmented encysted samples satisfy the smoothness evaluation in terms of the adaptive threshold (0.57). In contrast, we demonstrate that only 1\% of sensitive samples meet the adaptive smoothness threshold (LPIPS $=0.65$). For the time cost, generating one smooth sensitive sample takes more than one hour, which is infeasible in practice. Our PublicCheck only takes 9 seconds to generate one smooth encysted sample for fingerprinting. The visualization of the successful sensitive samples is shown in Figure~\ref{fig_ss} in the Appendix, which does not seem smooth for human perception despite having a small pixel-wise error level. 

\subsection{Evaluation of Adaptive Attacks}
\label{sec_evaluation_of_adaptive_attacks}

\subsubsection{Adaptive attack settings} 

The attacker's capability of collecting verification samples is limited as the verification samples are provided after the paywall (\eg, limited by the amount of API purchased). To demonstrate the worst-case, we consider the scenario when a large number of verification samples and associated labels are directly collected by the adversary, denoted by \texttt{LK-I}, mimicking an information leakage of the exact generative and verification capabilities. 
We assume the sampled data for the generative model is encrypted and transmitted only between the model developer and the independent secure verification server. Therefore, it is not possible for the adversary to gain access to the training data. We further consider the ramification if the sampled data was stolen as another worst-case scenario, \texttt{LK-II}. 

First, we consider that the attacker may collect a set of $N$ collected verification samples for each class along with original samples (\eg, public available samples from the same domain, or the sampled data to train the generative model), \ie, \texttt{LK-I}+\texttt{LK-II}. Three types of adaptive attacks (AA1-3) are evaluated under such a scenario. Furthermore, we consider two attacks (AA4-5) when the adversary leverages a substitute generative model trained on the \texttt{LK-II}.

\noindent \textit{\textbf{AA1: Adaptive detector.}} A threshold-based detection that uses leaked samples to identify original samples from verified samples. The detection criteria include prediction behavior (\eg, confidence) via a classifier, perceived similarity of images, and reconstruction error via an autoencoder.

\noindent \textit{\textbf{AA2: Adaptive classifier.}} A DNN-based classifier conducts binary classification on leaked samples. 

\noindent \textit{\textbf{AA3: Adaptive adversarial training.}} Leaked samples are used to conduct adversarial training on the attacked model (\eg, backdoored version) in order to bypass verification (produce the correct prediction on the verification sample).

\noindent \textit{\textbf{AA4: Substitute discriminator.~}} In the scenario \texttt{LK-II}, a substitute generative model associated with a discriminator (such as GAN or VAE-GAN) is trained to distinguish between generated images and original images.

\noindent \textit{\textbf{AA5: Substitute adversarial training.}} Using the substitute generative model trained in \texttt{LK-II}, the adversary simulates the encysted sample generation procedure to obtain verification samples for adversarial training. 

\subsubsection{Adaptive attacks evaluation and analysis}

To evaluate AA1, AA2, and AA3, $N$ encysted samples ($N = 10, 50, 1000, 2000$) for each class and 20\% original training data are assumed to be collected by the adversary.

\noindent \textit{\textbf{Results for AA1 and AA2.}}
We evaluate the prediction confidence values, the logit values before the softmax, and the Structural Similarity Index Measure (SSIM) perceived similarity values of encysted samples, compared with original samples. 
We find that the prediction confidence values and the logit values of encysted samples under different settings are similar in mean and distribution. 
We further evaluate the perceived similarity (SSIM) and Learned Perceptual Image Patch Similarity (LPIPS) values on generated encysted samples, under different settings ($N = 10, 50, 1000, 2000$). 
The experimental result shows that the similarity is high (mean SSIM 0.9, LPIPS 0.05) between the encysted samples from 20\% baseline and original samples. Even the similarity is dropped when decreasing the data size to 5\% (mean SSIM 0.85, LPIPS 0.06) and 10\% (mean SSIM 0.88, LPIPS 0.052), and similarity increases to 20\% baseline equivalent or better after applying the quality refiner strategies. 
To conclude, it is not practical to distinguish encysted samples from the original samples using the threshold in terms of prediction behaviors and perceived/perceptual similarity.

Additionally, we also examined the indistinguishability of the samples through a stronger detector based on reconstruction errors 
between the input image and its reconstructed image via a pre-trained autoencoder to differentiate the verification samples. The result shows that more than 91.2\% (out of 1,000) pixel-wise perturbations are identified by the reconstruction error-based detector, while the detection rate for our structural perturbation is less than 2.3\% (out of 1,000) using the same reconstruction error threshold and pre-trained autoencoder. As demonstrated in Figure~\ref{fig_reconstruction_quality} in the appendix, the pixel-wise perturbation results in a larger reconstruction error. A reconstruction error threshold (\eg, 10, under different versions of pre-trained autoencoder) could be used to identify most of the verification samples derived from existing pixel-wise perturbation based approaches. 
The distribution of the reconstruction error of encysted samples is within the value range for original samples. Namely, the identification of encysted samples is at the expense of misidentifying original samples. For example, the threshold used to identify 90\% encysted samples results in roughly a 50\% false positive rate on original samples. This reconfirms the indistinguishability of the encysted samples. 
Furthermore, the results of the AA2 confirm the indistinguishability of our encysted samples that the real/fake binary classification result for the classifier is equal to the random guess (50\%) under all data size settings. 
On the other hand, our smoothness filtering after encysted sample generation could also include the conditions incorporating these criteria to further ensure that verification samples are within the indistinguishable value range, based on the infinite generation capability. 
The demonstrated indistinguishability will force the service provider to give up verification manipulation, as the provider can neither identify a query nor treat all queries as verification queries (leading to the loss of service quality). 

\noindent \textit{\textbf{Results for AA3.}}
Given the target deployed model $TM$, the adaptive adversaries combine these verification samples and labels into the training data in order to construct the compressed model $TM^\prime$ (hard case for verification). Next, the verification function is used to generate new encysted samples to distinguish $TM^\prime$ from $TM$. 
We first test the performance of the Sensitive Sample against the AA3 detector. The results indicate that the detection accuracy sharply decreases when the network is trained with sensitive samples along with their clean pairs. Based on adversarial training with $N = 1000$, the attacked model could bypass more than 80\% of sensitive samples created by the baseline~\cite{he2019sensitive}, resulting in a similar performance to the randomly selected training samples. Adversarial training has successfully learned the distribution of fingerprinting patterns created by Sensitive Samples, demonstrating that AA3 is a strong adaptive attack. 
However, the detection accuracy of PublicCheck for $TM^\prime$ is not affected after applying adversarial training with different amounts of publicly released encysted samples (less than 2\% of verification samples are compromised by adaptive attack). We acknowledge that the abusive collection of verification samples shall be restricted by the prescribed amount of API consumed. 

The robustness against the strong adversarial training-based attack and model extraction is based on uncertainty and randomness in the generation of encysted samples. The released augmented encysted samples could be regarded as randomly sampled points distributed around uncertain areas of the decision boundary of the target model, which are treated as noisy samples and excluded from the learning process (see detailed explanation in Appendix~\ref{appendix_ability_against_adaptive_attacks}).

\noindent \textit{\textbf{Results for AA4 and AA5.}}
We also demonstrate the performances of AA4 and AA5 attacks when the adversary leverages a substitute generative model trained on the \texttt{LK-II}. 
The real/fake classification results of the discriminator of a GAN are equal to the random guess (50\%) under all data size settings. 
The discriminator of a VAE-GAN slightly outperforms classification (roughly 49.5\% across all data size settings), which will be reduced to a random guess level after incorporating the VAE-GAN into the autoencoder of PublicCheck training. 
The performance of the adversarial training, using the 2000 surrogate encysted samples via the substitute generative model trained on \texttt{LK-II}, is similar to that of AA3, without impacting the detection accuracy and bypassing less than 1\% of encysted samples. 
These results demonstrate the robustness of PublicCheck against adversarial attacks, especially after incorporating the adversarial strategies into the filter. We also note that discussions of Denial-of-Service attacks (\eg, abuse of the  generation of verification samples by compromised accounts) and attacks on communication protocols (\eg, session key distribution) are beyond the scope of this paper.

\section{Conclusion}
We propose a practical run-time deep models verification scheme PublicCheck, for ensuring the integrity of cloud-enabled deep models with public verifiability. 
We show that PublicCheck can withstand model integrity attacks and compression. 
We also show PublicCheck can achieve three important requirements that \textit{(i)} [Lightweight] only five randomly selected encysted fingerprinting samples are sufficient to achieve a 100\% verification and zero false-positive rates, without any extra coverage-guarantee mechanism; \textit{(ii)} [Smoothness] our structured perturbation in the latent space has reasonable controllability and disentanglement strategies; \textit{(iii)} [Anti-counterfeiting] linear perturbations in the latent space will result in non-linear structural change in the pixel space. These augmented encysted samples could be randomly scattered over the vicinity of the decision boundary of the target model and multi-class reference samples for random augmentation will further help prevent blind spots on the decision boundary. 
We also demonstrate the robustness of PublicCheck against five adaptive attacks. 
We hope that PublicCheck could be amenable to future-generation verification in deep neural networks. 

 \section*{Acknowledgement}
 
 The work has been supported by the Cyber Security Research Centre Limited whose activities are partially funded by the 1348 Australian Government’s Cooperative Research Centres Programme.

{
\bibliographystyle{IEEEtran}
\bibliography{ref}

\begin{thebibliography}{10}
\providecommand{\url}[1]{#1}
\csname url@samestyle\endcsname
\providecommand{\newblock}{\relax}
\providecommand{\bibinfo}[2]{#2}
\providecommand{\BIBentrySTDinterwordspacing}{\spaceskip=0pt\relax}
\providecommand{\BIBentryALTinterwordstretchfactor}{4}
\providecommand{\BIBentryALTinterwordspacing}{\spaceskip=\fontdimen2\font plus
\BIBentryALTinterwordstretchfactor\fontdimen3\font minus
  \fontdimen4\font\relax}
\providecommand{\BIBforeignlanguage}[2]{{%
\expandafter\ifx\csname l@#1\endcsname\relax
\typeout{** WARNING: IEEEtran.bst: No hyphenation pattern has been}%
\typeout{** loaded for the language `#1'. Using the pattern for}%
\typeout{** the default language instead.}%
\else
\language=\csname l@#1\endcsname
\fi
#2}}
\providecommand{\BIBdecl}{\relax}
\BIBdecl

\bibitem{li2021hidden}
S.~Li, H.~Liu, T.~Dong, B.~Z.~H. Zhao, M.~Xue, H.~Zhu, and J.~Lu, ``Hidden
  backdoors in human-centric language models,'' in \emph{Proceedings of the
  2021 ACM SIGSAC Conference on Computer and Communications Security}, 2021,
  pp. 3123--3140.

\bibitem{li2020invisible}
S.~Li, M.~Xue, B.~Z.~H. Zhao, H.~Zhu, and X.~Zhang, ``Invisible backdoor
  attacks on deep neural networks via steganography and regularization,''
  \emph{IEEE Transactions on Dependable and Secure Computing}, vol.~18, no.~5,
  pp. 2088--2105, 2020.

\bibitem{zhong2020backdoor}
H.~Zhong, C.~Liao, A.~C. Squicciarini, S.~Zhu, and D.~Miller, ``Backdoor
  embedding in convolutional neural network models via invisible
  perturbation,'' in \emph{Proceedings of the Tenth ACM Conference on Data and
  Application Security and Privacy}, 2020, pp. 97--108.

\bibitem{chen2017targeted}
X.~Chen, C.~Liu, B.~Li, K.~Lu, and D.~Song, ``Targeted backdoor attacks on deep
  learning systems using data poisoning,'' \emph{arXiv preprint
  arXiv:1712.05526}, 2017.

\bibitem{turner2018clean}
A.~Turner, D.~Tsipras, and A.~Madry, ``Clean-label backdoor attacks,'' 2018.

\bibitem{liu2017trojaning}
Y.~Liu, S.~Ma, Y.~Aafer, W.-C. Lee, J.~Zhai, W.~Wang, and X.~Zhang, ``Trojaning
  attack on neural networks,'' \emph{Network and Distributed Systems Security
  (NDSS) Symposium}, 2018.

\bibitem{ma2023beatrix}
W.~Ma, D.~Wang, R.~Sun, M.~Xue, S.~Wen, and Y.~Xiang, ``The {``Beatrix''}
  resurrections: Robust backdoor detection via {Gram} matrices,'' \emph{Network
  and Distributed System Security (NDSS) Symposium}, 2023.

\bibitem{he2019sensitive}
Z.~He, T.~Zhang, and R.~Lee, ``Sensitive-sample fingerprinting of deep neural
  networks,'' in \emph{Proceedings of the IEEE/CVF Conference on Computer
  Vision and Pattern Recognition}, 2019, pp. 4729--4737.

\bibitem{ma2021quantization}
H.~Ma, H.~Qiu, Y.~Gao, Z.~Zhang, A.~Abuadbba, A.~Fu, S.~Al-Sarawi, and
  D.~Abbott, ``Quantization backdoors to deep learning models,'' \emph{arXiv
  preprint arXiv:2108.09187}, 2021.

\bibitem{tian2022stealthy}
Y.~Tian, F.~Suya, F.~Xu, and D.~Evans, ``Stealthy backdoors as compression
  artifacts,'' \emph{IEEE Transactions on Information Forensics and Security},
  vol.~17, pp. 1372--1387, 2022.

\bibitem{adi2018turning}
Y.~Adi, C.~Baum, M.~Cisse, B.~Pinkas, and J.~Keshet, ``Turning your weakness
  into a strength: Watermarking deep neural networks by backdooring,'' in
  \emph{27th USENIX Security Symposium (USENIX Security 18)}, 2018, pp.
  1615--1631.

\bibitem{le2020adversarial}
E.~Le~Merrer, P.~Perez, and G.~Tr{\'e}dan, ``Adversarial frontier stitching for
  remote neural network watermarking,'' \emph{Neural Computing and
  Applications}, vol.~32, no.~13, pp. 9233--9244, 2020.

\bibitem{lukas2021deep}
N.~Lukas, Y.~Zhang, and F.~Kerschbaum, ``Deep neural network fingerprinting by
  conferrable adversarial examples,'' in \emph{International Conference on
  Learning Representations}, 2021.

\bibitem{fridrich2000robust}
J.~Fridrich and M.~Goljan, ``Robust hash functions for digital watermarking,''
  in \emph{Proceedings International Conference on Information Technology:
  Coding and Computing (Cat. No. PR00540)}.\hskip 1em plus 0.5em minus
  0.4em\relax IEEE, 2000, pp. 178--183.

\bibitem{kim2018disentangling}
H.~Kim and A.~Mnih, ``Disentangling by factorising,'' in \emph{International
  Conference on Machine Learning}.\hskip 1em plus 0.5em minus 0.4em\relax PMLR,
  2018, pp. 2649--2658.

\bibitem{burgess2018understanding}
C.~P. Burgess, I.~Higgins, A.~Pal, L.~Matthey, N.~Watters, G.~Desjardins, and
  A.~Lerchner, ``Understanding disentangling in $beta$-vae,'' \emph{arXiv
  preprint arXiv:1804.03599}, 2018.

\bibitem{zhang2018unreasonable}
R.~Zhang, P.~Isola, A.~A. Efros, E.~Shechtman, and O.~Wang, ``The unreasonable
  effectiveness of deep features as a perceptual metric,'' in \emph{Proceedings
  of the IEEE Conference on Computer Vision and Pattern Recognition}, 2018, pp.
  586--595.

\bibitem{razavi2019generating}
A.~Razavi, A.~van~den Oord, and O.~Vinyals, ``Generating diverse high-fidelity
  images with vq-vae-2,'' in \emph{Advances in Neural Information Processing
  Systems}, 2019, pp. 14\,866--14\,876.

\bibitem{van2017neural}
A.~Van Den~Oord, O.~Vinyals \emph{et~al.}, ``Neural discrete representation
  learning,'' in \emph{Advances in Neural Information Processing Systems},
  2017.

\bibitem{carlini2017towards}
N.~Carlini and D.~Wagner, ``Towards evaluating the robustness of neural
  networks,'' in \emph{2017 IEEE Symposium on Security and Privacy (SP)}.\hskip
  1em plus 0.5em minus 0.4em\relax IEEE, 2017, pp. 39--57.

\bibitem{li2019nattack}
Y.~Li, L.~Li, L.~Wang, T.~Zhang, and B.~Gong, ``Nattack: Learning the
  distributions of adversarial examples for an improved black-box attack on
  deep neural networks,'' in \emph{International Conference on Machine
  Learning}.\hskip 1em plus 0.5em minus 0.4em\relax PMLR, 2019, pp. 3866--3876.

\bibitem{wierstra2014natural}
D.~Wierstra, T.~Schaul, T.~Glasmachers, Y.~Sun, J.~Peters, and J.~Schmidhuber,
  ``Natural evolution strategies,'' \emph{The Journal of Machine Learning
  Research}, vol.~15, no.~1, pp. 949--980, 2014.

\bibitem{mohaghegh2020advflow}
H.~Mohaghegh~Dolatabadi, S.~Erfani, and C.~Leckie, ``Advflow: Inconspicuous
  black-box adversarial attacks using normalizing flows,'' \emph{Advances in
  Neural Information Processing Systems}, vol.~33, pp. 15\,871--15\,884, 2020.

\bibitem{lecun1998gradient}
Y.~LeCun, L.~Bottou, Y.~Bengio, P.~Haffner \emph{et~al.}, ``Gradient-based
  learning applied to document recognition,'' \emph{Proceedings of the IEEE},
  vol.~86, no.~11, pp. 2278--2324, 1998.

\bibitem{krizhevsky2009learning}
A.~Krizhevsky, G.~Hinton \emph{et~al.}, ``Learning multiple layers of features
  from tiny images,'' 2009.

\bibitem{karras2019style}
T.~Karras, S.~Laine, and T.~Aila, ``A style-based generator architecture for
  generative adversarial networks,'' in \emph{Proceedings of the IEEE
  Conference on Computer Vision and Pattern Recognition}, 2019, pp. 4401--4410.

\bibitem{karras2020analyzing}
T.~Karras, S.~Laine, M.~Aittala, J.~Hellsten, J.~Lehtinen, and T.~Aila,
  ``Analyzing and improving the image quality of stylegan,'' in
  \emph{Proceedings of the IEEE/CVF conference on computer vision and pattern
  recognition}, 2020, pp. 8110--8119.

\bibitem{he2016deep}
K.~He, X.~Zhang, S.~Ren, and J.~Sun, ``Deep residual learning for image
  recognition,'' in \emph{Proceedings of the IEEE conference on computer vision
  and pattern recognition}, 2016, pp. 770--778.

\bibitem{simonyan2014very}
K.~Simonyan and A.~Zisserman, ``Very deep convolutional networks for
  large-scale image recognition,'' \emph{arXiv preprint arXiv:1409.1556}, 2014.

\bibitem{pang2020trojanzoo}
R.~Pang, Z.~Zhang, X.~Gao, Z.~Xi, S.~Ji, P.~Cheng, and T.~Wang, ``Trojanzoo:
  Everything you ever wanted to know about neural backdoors (but were afraid to
  ask),'' in \emph{arXiv Preprint}, 2020.

\bibitem{gu2017badnets}
T.~Gu, B.~Dolan-Gavitt, and S.~Garg, ``Badnets: Identifying vulnerabilities in
  the machine learning model supply chain,'' \emph{arXiv preprint
  arXiv:1708.06733}, 2017.

\bibitem{wang2021zero}
Z.~Wang, ``Zero-shot knowledge distillation from a decision-based black-box
  model,'' in \emph{International Conference on Machine Learning}.\hskip 1em
  plus 0.5em minus 0.4em\relax PMLR, 2021, pp. 10\,675--10\,685.

\end{thebibliography}
}
\appendices

\begin{figure}[b]
\centering  
\includegraphics[width=0.9\linewidth]{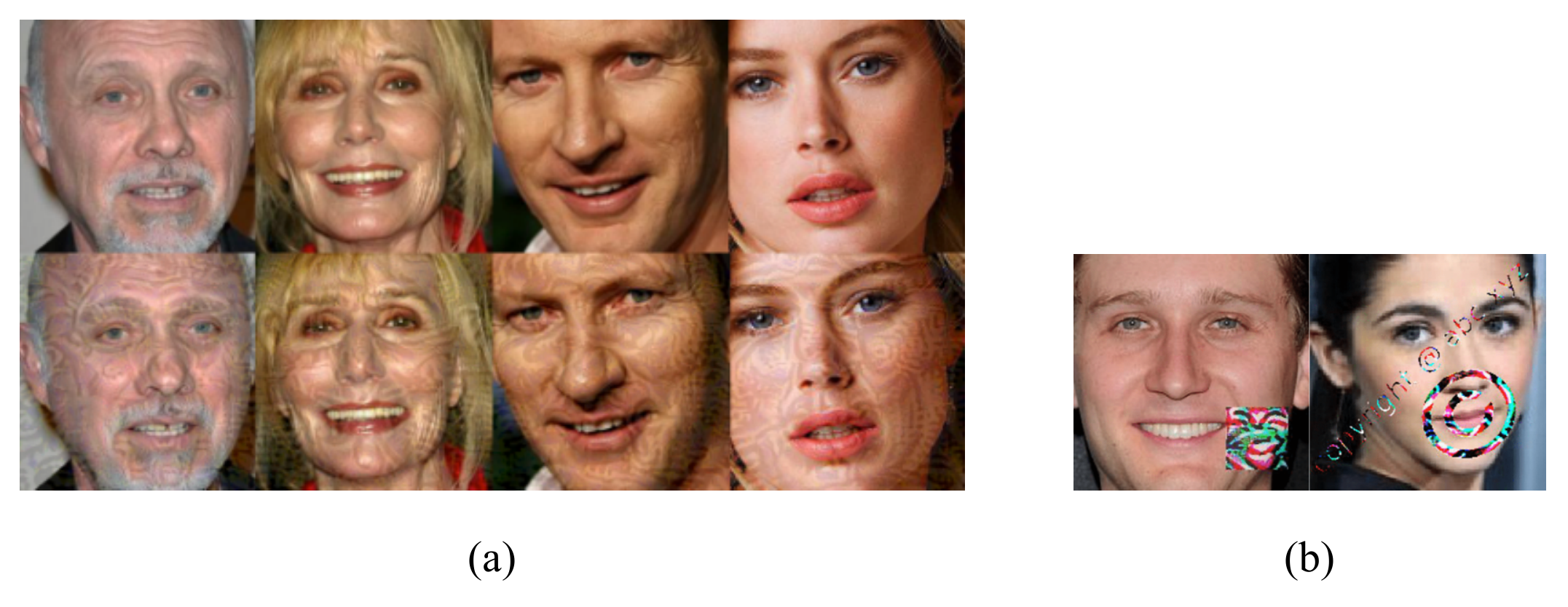}
\caption{A verification sample is generally created from an original example by adding a trigger or pixel perturbation in existing works. Such pixel-level patterns can be distinguished by human perception. (a) Sensitive Samples~\cite{he2019sensitive} for private verification of model integrity (Second row). The first row is the original images. (b) Triggers (Trojan Square and Trojan Watermark) used in Backdoor attacks~\cite{liu2017trojaning}.}
\label{fig_ss}
\end{figure}

\section{Autoencoder-based Generative Models}
\label{appendix_generative_models}
\subsection{Variational autoencoder (VAE)}
\label{appendix_vae}
The loss of VAE incorporated with a TC term is given as follows:
\begin{equation}
\scriptsize
\begin{aligned}
\mathcal{L}_{VAE}=\mathcal{L}_{Rec}- D_{KL}[(q_{\phi } (z|x)||p(z))] - \upsilon ~TC(z)
\end{aligned}
\label{eq_vqlosssim}
\end{equation}
The first term $\mathcal{L}_{Rec}$ is the reconstruction error based on LPIPS, which assesses whether the latent codes $z$ are informative enough to reassemble the original instance.
The second part is a regularization term, to push Encoder $ q_{\phi } (z|x)$ to match the prior distribution $p(z)$, \eg, a Gaussian distribution.
The third part is the TC term, to measure the dependence for multiple random variables, with $\upsilon=40$. We replace the default pixel-wise reconstruction evaluation with the perceptual evaluation metric Learned Perceptual Image Patch Similarity (LPIPS)~\cite{zhang2018unreasonable}, calculated as a weighted difference between two VGG16 embeddings. 

\subsection{VQ-based Generative Models}
\label{appendix_vq}

\begin{figure}[t]
\centering  
\includegraphics[width=0.9\linewidth]{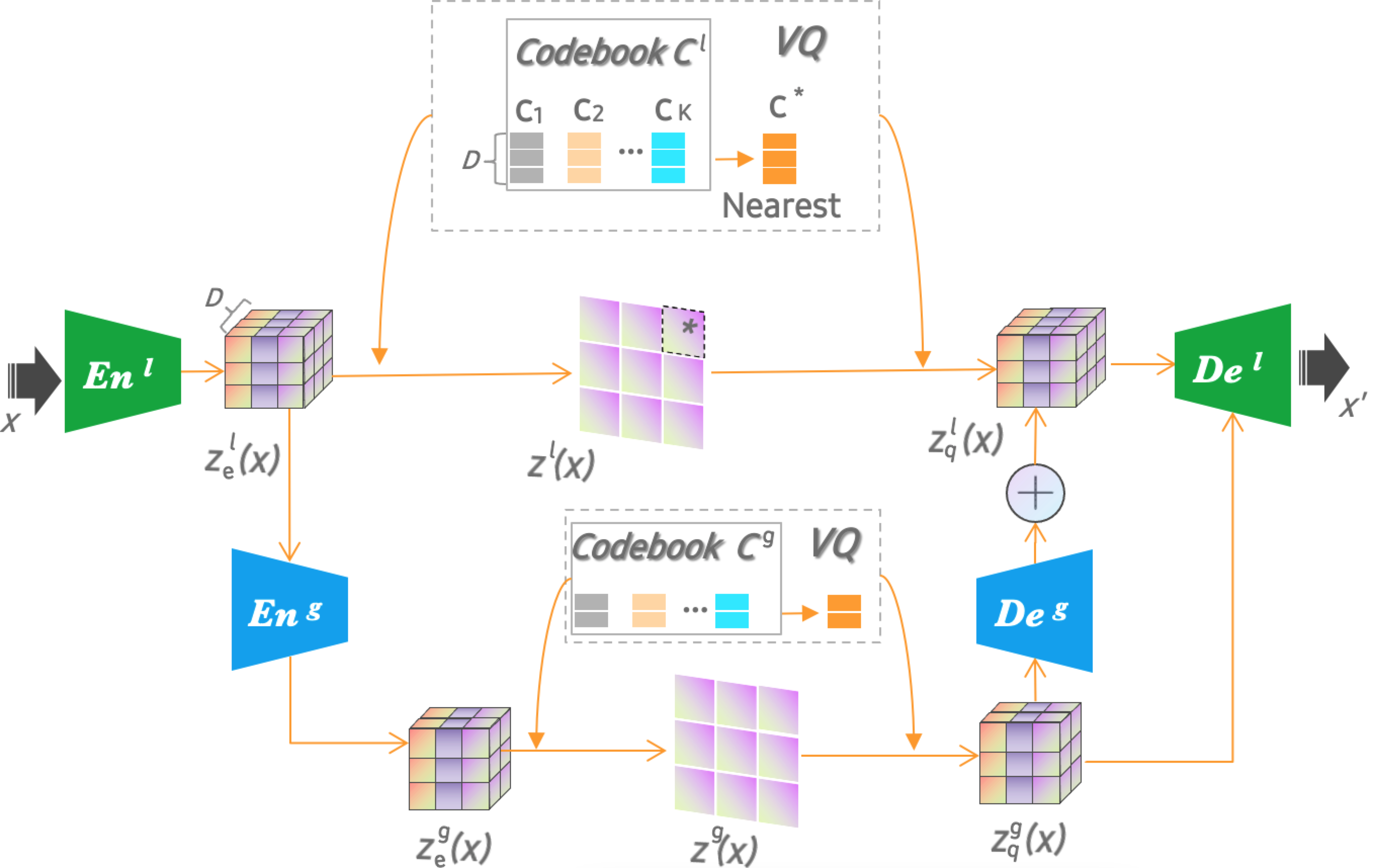}
\caption{Demonstration of a VQ-based generative model.}
\label{fig_vq}
\end{figure}

The codebook $C$ defines the commonly-shared latent embedding space $C\in R^{K\times D}$,  consisting of $K$ categorical embedding items with D dimension, \ie, $C_i\in R^D, ~i\in \{1,2,\ldots,K\}$. 
The encoder is a non-linear mapping from the input instance $x$ in the pixel space the latent representation $z_{e}(x)\in R^{W\times H\times D}$, namely, $W\times D$ latent embedding vectors with $D$ dimension ($z^{(i,j)}_{e}\in R^D, ~i\in \{1,2,\ldots,W\},~j\in \{1,2,\ldots,H\}$). Then, the latent representation $z_{e}(x)$ is further mapped to a discrete latent matrix $z^{(i,j)}\in R^1$. Here, each $z^{(i,j)}$ is the index of the nearest embedding items $C_{nrs}$ in the codebook for each $z^{(i,j)}_{e}$ via nearest neighbour searching $\argmin_m\left \| z^{(i,j)}_e(x)-C_m \right \|$, also called Vector Quantized, as demonstrated in Figure \ref{fig_vq}. 
The decoder reconstructs back to pixel space using the queried embedding items $z_q(z)$ corresponding to the discrete latent index matrix via another non-linear function. 
Trainable parameters for the model are the union of parameters of the encoder, decoder, and codebook.
The loss function is:
\begin{equation} \scriptsize
\begin{aligned}
Dist(\mathbf{x}-De(\mathbf{z_q}))+\| sg[En(\mathbf{x})]-\mathbf{C_{nrs}}\|_{2}^{2}
+\beta \| sg[\mathbf{C_{nrs}}]-En(\mathbf{x})\|_{2}^{2}
\end{aligned}
\label{eq_vqlossreal}
\end{equation}
The operator $sg$ refers to a stop-gradient operation that blocks gradients from flowing into its argument, and $\beta$ is a hyperparameter to control the reluctance to change the code corresponding to the encoder output. 
The first term is the gradient of the reconstruction error $Dist(\mathbf{x}-De(\mathbf{z_q}))$, which will be back-propagated through the decoder, and to the encoder using the straight-through gradient estimator. 
The procedures of training encoder and decoder, and exponential moving average updating of the codebook are followed \cite{van2017neural,razavi2019generating}, while replacing the default pixel-wise reconstruction evaluation with the perceptual evaluation metric Learned Perceptual Image Patch Similarity (LPIPS) \cite{zhang2018unreasonable}. 

Although the training time of the VQ generative model takes hours for 256x images, that is a once-off procedure on the server side that could be carried out in parallel across several GPUs or by using the pre-trained model for transfer learning to further reduce the training time. Additionally, the generative could be reused for similar datasets, \eg, the model pre-trained on the FFHQ face dataset could be used for other human face datasets.

\section{Boosting Strategies for Infinite Supply of Verification Samples}
\label{appendix_booster}
One of the advantages of our approach is that it enables large amounts of disposable verification samples. 
Firstly, for each user, only 5 verification samples can achieve a 100\% verification accuracy (Section 5.3).
Secondly, an infinite number of perturbations (within a specific noise range) could be sampled from the latent space, producing a plethora of encysted samples (1 code-1 reference). A set of reference original samples would further enhance the number of verification samples (1 code $n$ reference). 
The amount could still be further boosted to infinity by combining existing latent codes (\eg, selecting two or more latent codes simultaneously to add the same or varying perturbations, n code strategy), or extending the dimension of the latent representation (extended dimension code strategy), or assuming different noise distributions (extended distribution-code strategy).
Due to the strong and efficient generation ability of our fingerprinting approach, it is possible to provide the once-off verification in practice. 
Further, the attacker's capability of collecting verification samples is limited as the verification is provided with API purchasing and under a verification budget (\eg 10 times per day per client).

\section{Black-box Settings}
\label{appendix_black_box_settings}
We redefine the term black-box in our work into twofold: \textit{(i)} there is only black-box access to the model to perform verification; \textit{(ii)}  the design of the fingerprinting of the model is conducted with black-box knowledge about the model. 
The existing pixel perturbation-based fingerprinting approaches require white-box knowledge about the certified model, which is not always available to the participants in the MLaaS life cycle, such as the model agents. 
The licensed model obtained by the model agent is typically protected or encapsulated as executable files. 
Additionally, model files may be encrypted by compilation, shelling, or confusion, making them difficult to decompose. 
Furthermore, a model may not be gradient-based, such as simulated annealing or evolutionary strategies. In such circumstances, verification samples can only be generated in a black-box fashion. 
Moreover, with the white-box knowledge restriction, any users who do not have white-box access to the model will not be able to utilize the model fingerprinting services. 
Our approach eliminates the white-box knowledge restriction in favor of a more practical black-box knowledge scenario. 
The generalized and practical black-box knowledge assumption on the model brings more benefits, such as the provision for the infringement assessment of two deployed models by third parties, or the provision of regular integrity checks remotely from the end users' side rather than the server's.

\section{Datasets and Settings}
\label{appendix_datasets_and_settings}
\noindent \textit{(i)} MNIST consists of 28 $\times$ 28 grayscale handwritten digits 0-9, and has a training set of 55,000 instances and a test set of 10,000 instances. 

\noindent \textit{(ii)} The CIFAR-10 dataset consists of 60,000 32$\times$32 color images in 10 classes, with 6,000 images per class.

\noindent \textit{(iii)} CIFAR-100 dataset contains 100 classes with 600 images per class. The images are resized to 64px in our experiments. 

\noindent \textit{(iv)} Flickr-Faces-HQ (FFHQ) consists of 70,000 high-quality 1,024$\times$1,024 resolution human face images from Flickr and is associated with considerable variation in terms of age, ethnicity, and image background. The images are resized to 256px in our experiments.

Table~\ref{tab_datasets} shows the classification models used in our experiments for each of these datasets, and the corresponding model accuracy. 

\begin{table}[t]
\scriptsize
\centering
\caption{Classification model and baseline accuracy.}\label{tab_datasets}
\resizebox{0.8\linewidth}{!}{
\begin{tabular}{llll}
\toprule
\multicolumn{1}{c}{\textbf{Dataset}} &
  \multicolumn{1}{c}{\textbf{Task}} &
  \multicolumn{1}{c}{\textbf{Model}} &
  \multicolumn{1}{c}{\textbf{Accuracy}}\\  
  \hline
MNIST    & Digit Classification  & LeNet    & 99.97\\
CIFAR-10 & Image Classification  & ResNet18~\cite{he2016deep} & 93.59\\
FFHQ     & Gender Classification & VGG16~\cite{simonyan2014very}    & 89.87\\ \bottomrule
\end{tabular}}
\end{table} 

\begin{table}[t]
\centering
\scriptsize
\caption{Default hyperparameter settings.}\label{tab_hyperparameter}
\resizebox{0.75\linewidth}{!}{
\begin{tabular}{@{}llll@{}}
\toprule
\multirow{2}{*}{\textbf{Parameter}} & \multicolumn{3}{c}{\textbf{Default Values}} \\ \cmidrule(l){2-4} 
 & \textbf{MNIST} & \textbf{CIFAR-10} & \textbf{FFHQ} \\ \midrule
\textbf{Input size} & 28 & 256 & 256 \\
\textbf{$\beta$} & N/A & 0.25 & 0.25 \\
\textbf{Batch size} & 16 & 128 & 128 \\
\textbf{Codebook size} & N/A & (512,64) & \begin{tabular}[c]{@{}l@{}}bottom (512,64)\\ top (256,32)\end{tabular} \\
\textbf{Training steps} & 50 & 25000 & 25000 \\
\textbf{Learning rates} & 1e-4 & 2e-4 & 2e-4 \\ \bottomrule
\end{tabular}
}
\vspace{5mm}
\end{table}

Our DNN techniques are implemented using PyTorch, backdoor attacks are deployed in the TrojanZoo~\cite{pang2020trojanzoo} platform, and the PublicCheck in Python. 
The number of instances that are randomly sampled as reference samples is generally 3-4 times of the required number of the encysted sample for fingerprinting. For example, if four encysted samples for fingerprinting are desired, accordingly, we randomly select 20 reference samples to conduct augmentation, followed by randomly selecting 4 augmented encysted samples that satisfied the smooth appearance. 
Each entry in the results of performance for fingerprinting is averaged over 1,000 repetitions of augmentation and selection. 

\section{Quality Refiner Strategies}
\label{booster}
The quality refiner consists of data augmentation and knowledge distillation from the target model.
One method for loosening the restrictions regarding the amount of training data accessible for training generative models is data augmentation, such as Differentiable Augmentation (DiffAugment), in which the performance of a generative model trained on 100 seed images for augmentation is comparable to that of the model trained on the entire dataset (demonstrated in the human face, CIFAR-10 and Imagenet).  
On the other hand, during autoencoder training, there is also auxiliary knowledge that we can draw upon, namely, the well-trained classifier trained on the entire training dataset, even with only black-box access. 
Therefore, the decision-based black-box knowledge distillation \cite{wang2021zero} is used to transfer the knowledge of the well-trained classifier (black-box teacher) to the student classifier that consists of the encoder portion (and the common codebook) + one fully connected layer and softmax, using the augmented samples (\eg, rotation between $[-10^{\circ}, 10^{\circ}]$  ) and associated labels. 
The soft label of each training sample can be constructed with the value of sample robustness (the distance from a sample to the targeted decision boundary) and used for training the student via knowledge distillation. 
Then, we train the decoder portion of the network (while fine-tuning the encoder weights with a smaller learning rate) using the augmented samples.

\section{Architectures and Hyperparameters}
\label{appendix_architectures_and_hyperparameters}
The VAE-based generative model (attribute-level disentanglement) is used for MNIST, 
and the architecture and hyperparameters are similar to the settings in \cite{kim2018disentangling}. The encoder consists of four convolutional layers with steps of 2 and a window size of 4 $\times$ 4, followed by two fully connected layers with 128, 2 $\times$ 10 hidden units respectively. In the decoder, there is one fully connected layer (128 hidden units), followed by four transposed convolutions with stride 2 and window size 4 $\times$ 4. 
The VQ-based generative model (abstraction-level disentanglement) is used for CIFAR and FFHQ.
As the CIFAR-10 images are of low fidelity, we combine the top and bottom encoder into one,  \ie, the ordinary VQ-based autoencoder in Appendix~\ref{appendix_vq} is used. 
The architecture and hyperparameters of the ordinary VQ-based autoencoder are similar to the settings in~\cite{van2017neural} and that of the two-encoder VQ-based generative model (abstraction-level disentanglement) for FFHQ are similar to the settings in~\cite{razavi2019generating}. 
The encoder consists of two convolutional layers with a stride of 2 and a window size of 4 $\times$ 4, followed by two residual 3 $\times$ 3 blocks (implemented as ReLU, $3 \times 3$, ReLU, 1 $\times$ 1), all of which have 128 hidden units. In the decoder, there are two residual 3 $\times$ 3 blocks, followed by two transposed convolutions with stride 2 and window size 4 $\times$ 4. 
Default hyperparameters are in Table~\ref{tab_hyperparameter}.

\begin{figure}[t]
\centering
\includegraphics[width=0.95\linewidth]{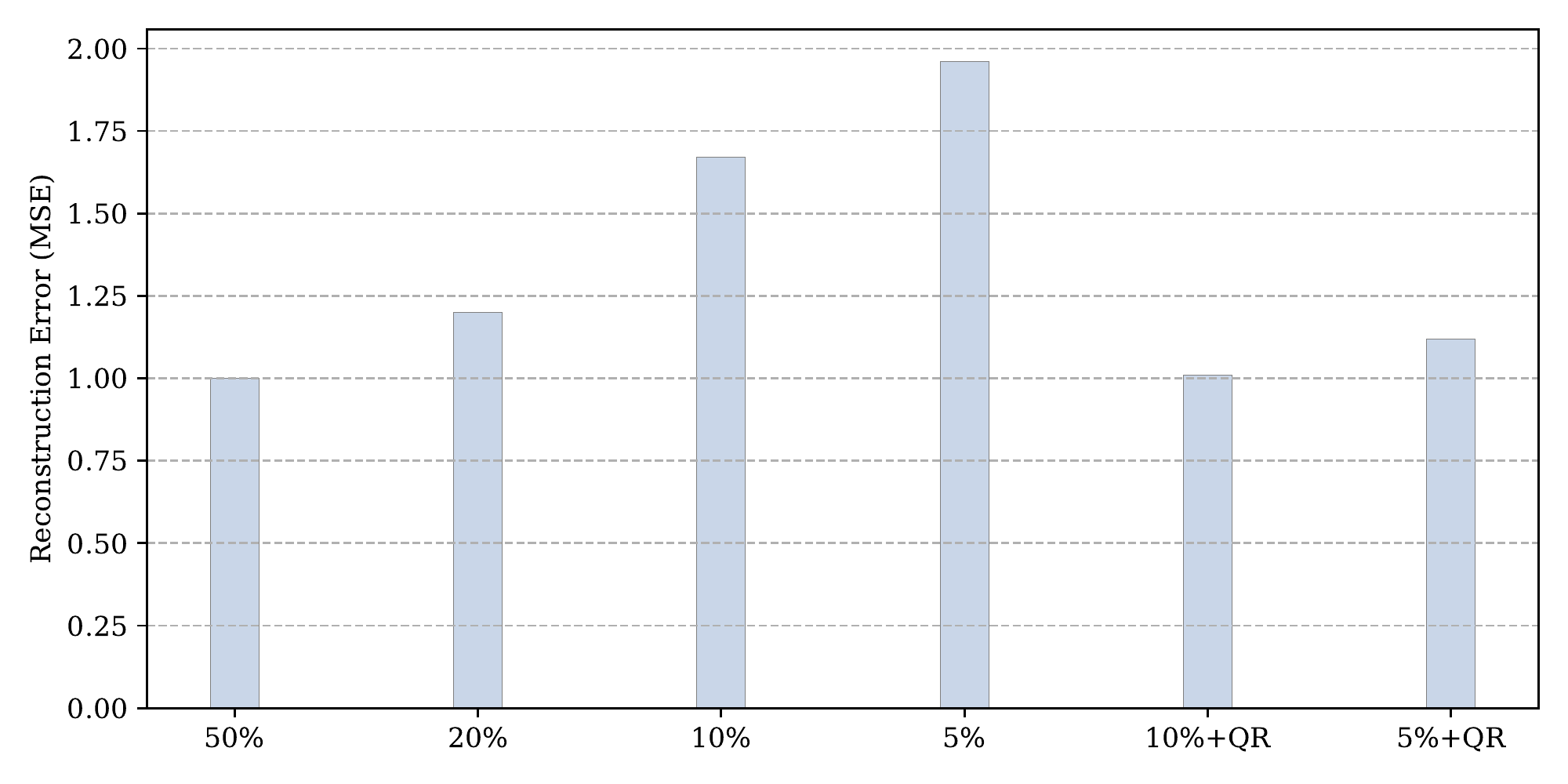}
\caption{MSE under various limited data settings.}
\label{appendix_mse}
\vspace{2mm}
\end{figure}

\begin{figure}[t]
\centering
\includegraphics[width=0.9\linewidth]{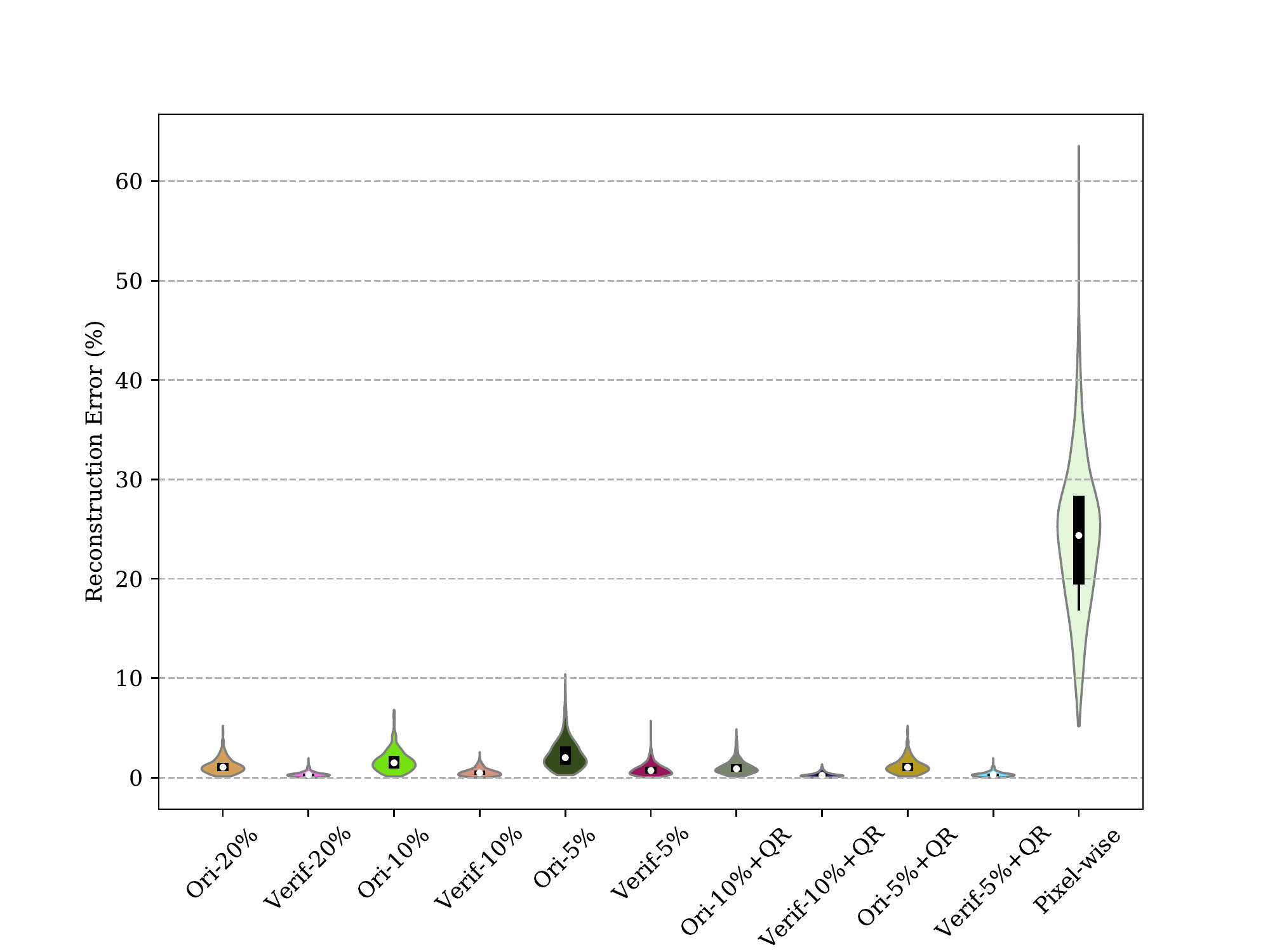}
\caption{Reconstruction errors for verification samples.}
\label{appendix_quality}
\vspace{2mm}
\end{figure}

\begin{figure}[t]
\centering 
\includegraphics[width=0.9\linewidth]{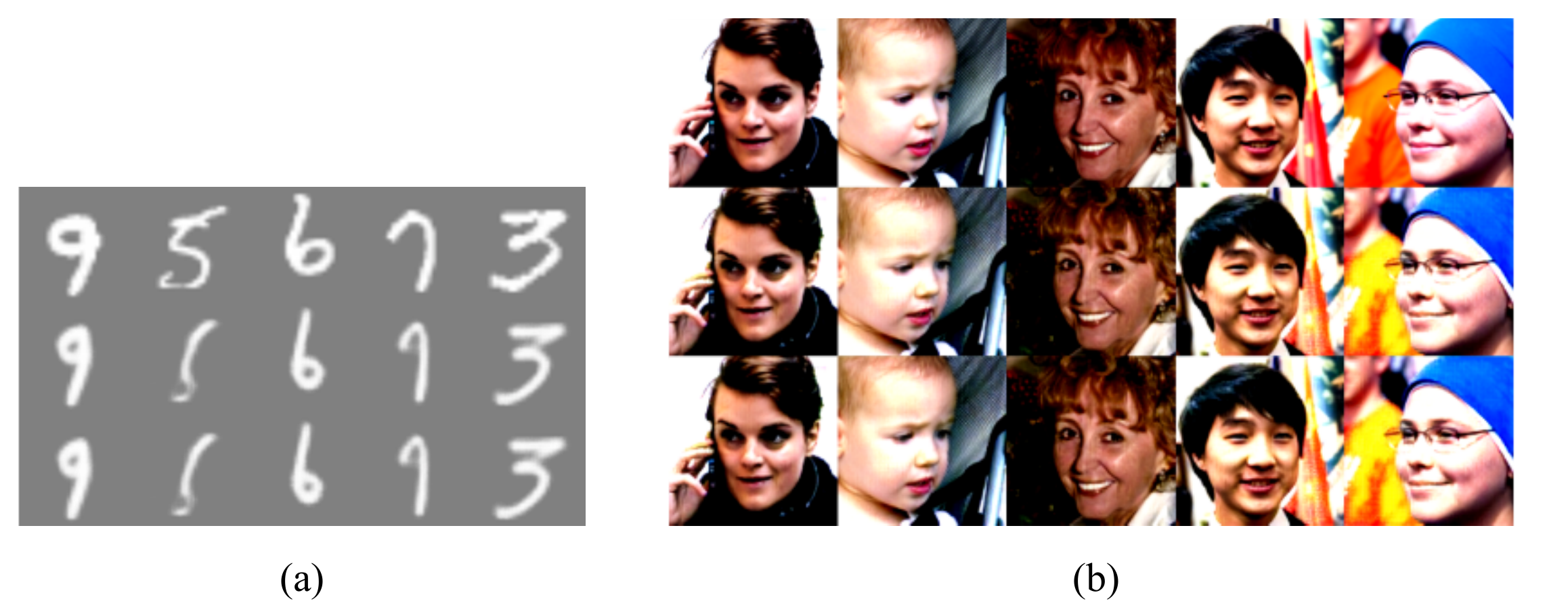}
\caption{Encysted samples for fingerprinting on MNIST (a) and FFHQ (b).  The first row is the original reference samples, second and last row is the inner (same prediction with the reference) and outer (different prediction) encysted samples, respectively, as close to the decision boundary.}
\label{fig_encysted_samples}
\vspace{2mm}
\end{figure}

\begin{table}[t]
\centering
\caption{Attack default settings.}
\resizebox{\linewidth}{!}{
\resizebox{0.8\linewidth}{!}{
\begin{tabular}{lp{5cm}p{2.5cm}}
\toprule
\multicolumn{1}{c}{\bf Attack} & \multicolumn{1}{c}{\bf Parameters} & \multicolumn{1}{c}{\bf Values} \\ \midrule
Training & learning rate, retrain epoch, optimizer, momentum, weight decay & 0.01, 50, SGD, 0.9, $2\times10^{-4}$ \\ \midrule
BadNet & poison\_percent & 0.1 \\ \midrule
TrojanNN & layer, neuron, optimizer,  lr, iter, threshold, target & logits, 2, PGD, 0.015, 20, 5,10 \\ \midrule
Clean-Label & poison generation, tau, epsilon, noise\_dim, iter & PGD, 0.4, 0.1, 100, 1000 \\ \bottomrule
\end{tabular}}
}
\label{tab_attack_settings}
\end{table}

\begin{figure}[t]
\centering
\includegraphics[width=0.95\linewidth]{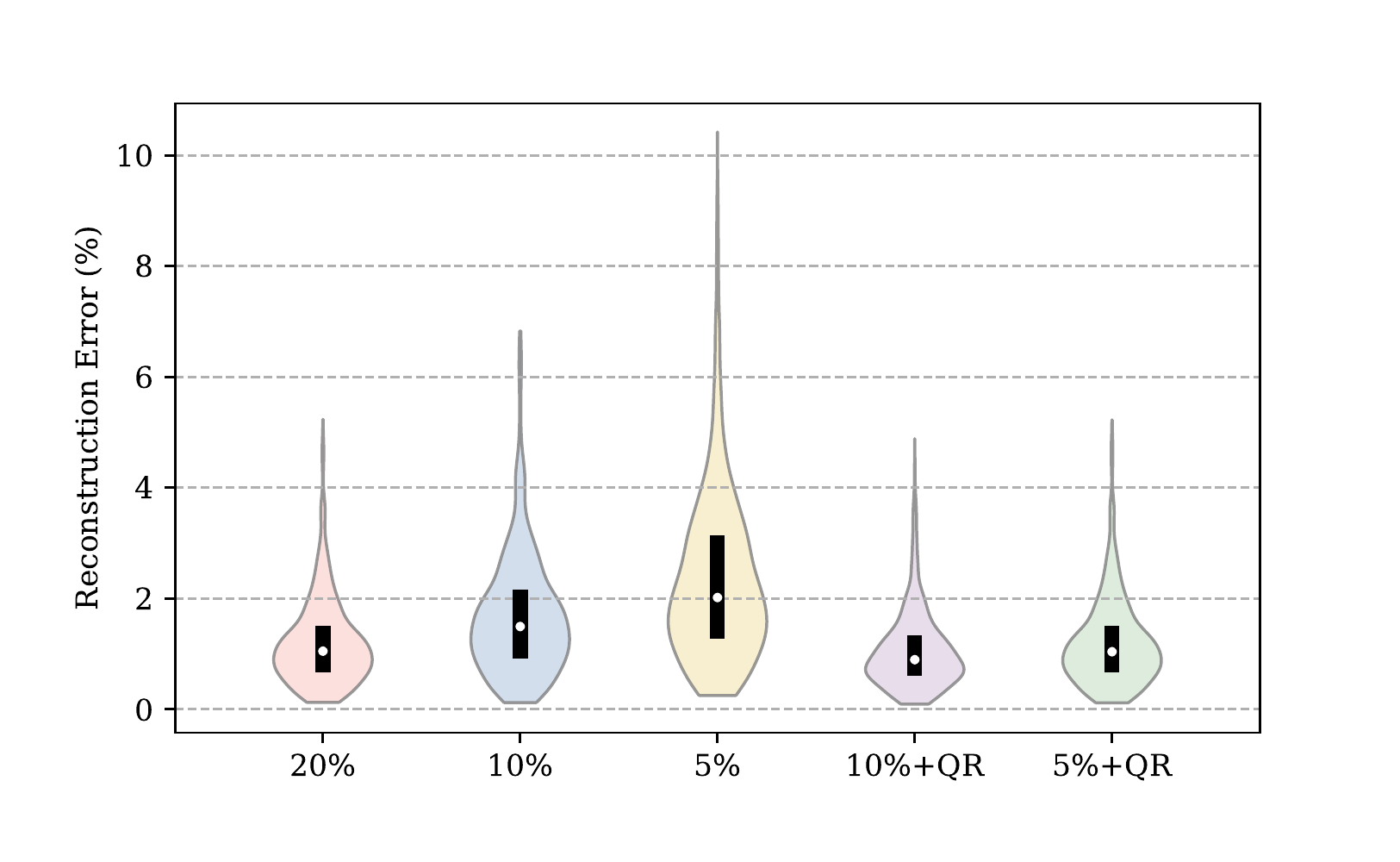}
\caption{Reconstruction quality of encysted samples under limited data settings.}
\label{fig_reconstruction_quality}
\vspace{3mm}
\end{figure}

\begin{figure}[t]
\centering  
\subfigure[]{\includegraphics[width=2.2in]{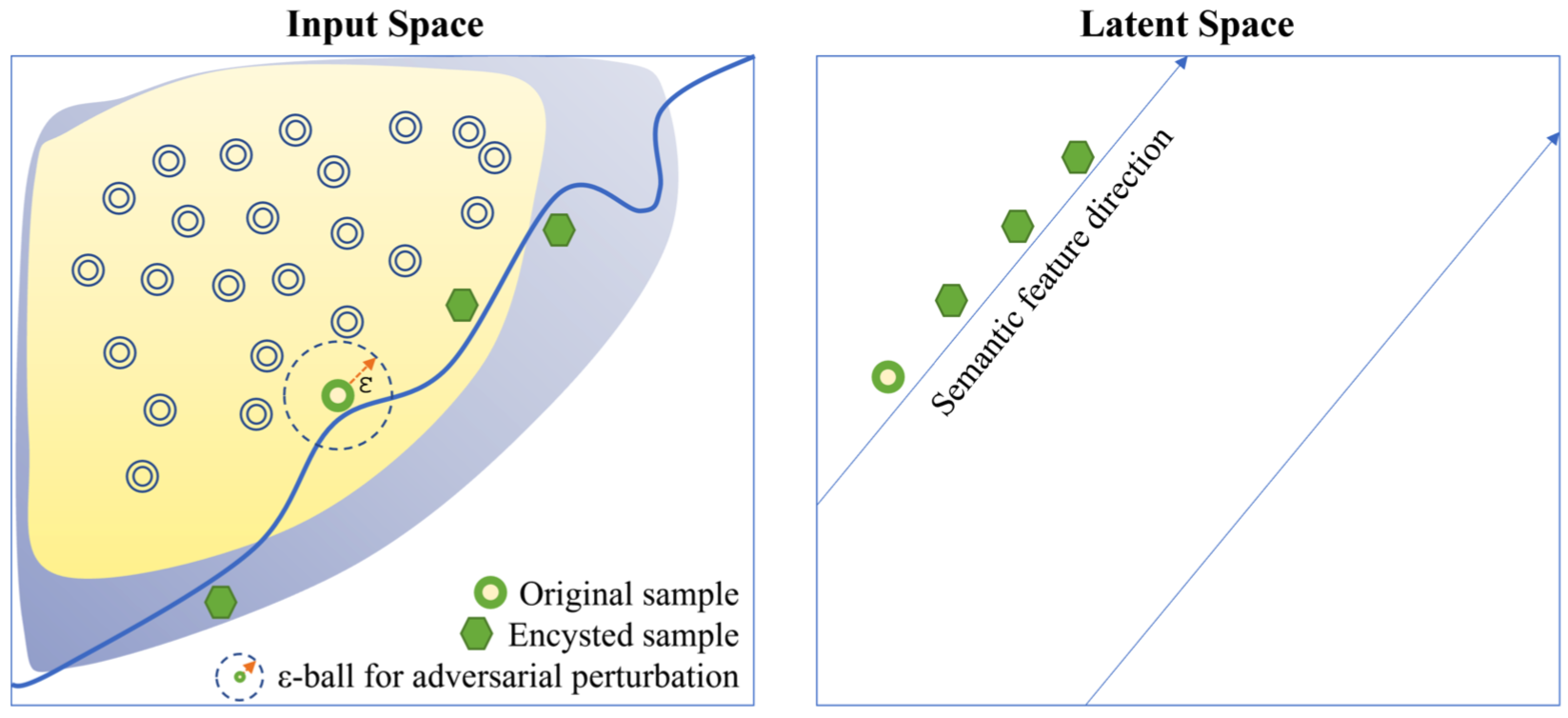}}
\subfigure[]{\includegraphics[width=1.2in]{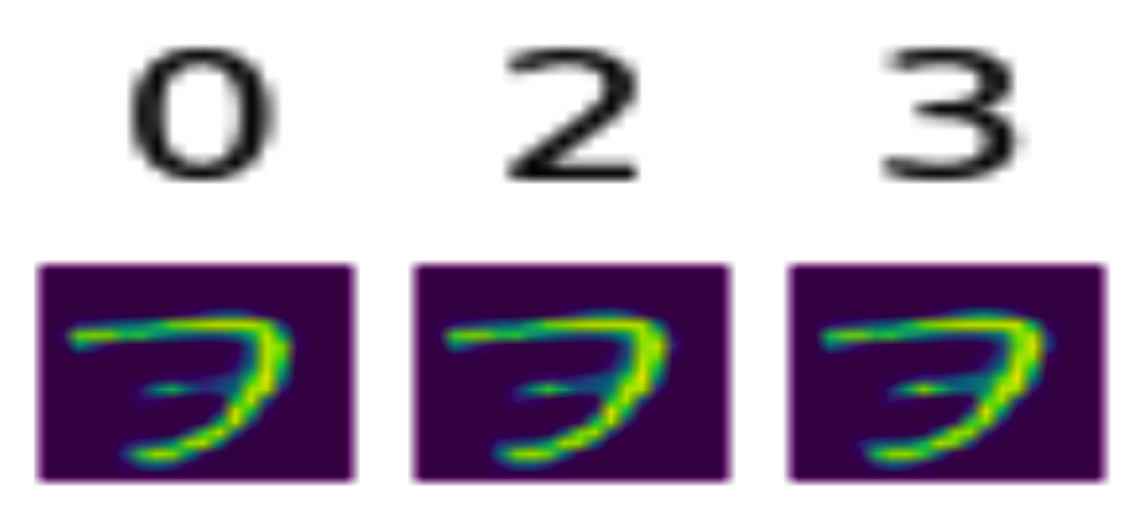}}
\caption{Demonstration of the robustness against adaptive attacks: (a) uncertainty of encysted samples; (b) the misclassification labels for three of the reconstructed images with similar latent perturbations are different.}
\label{fig_aar}
\vspace{4mm}
\end{figure}

\section{Ability against Adaptive Attacks}
\label{appendix_ability_against_adaptive_attacks}
Our approach could defeat such strong adaptive attacks for the following reasons. 
As shown in Figure~\ref{fig_aar}, adversarial perturbation in the pixel space is around an $\epsilon$-ball, which exhibits spatial clustering patterns. In addition, due to the restriction into a certain region of the decision boundary, the misclassification behavior caused by such pixel perturbation is also with some certainty. An adaptive attack may therefore use adversarial training to learn such patterns to fix a certain region of the decision boundary and defeat the pixel perturbation.

By contrast, our perturbation occurs  in the latent space after non-linear encoding and is then reconstructed into the pixel space through non-linear decoding. During each augmentation, these augmented encysted samples may be distributed over multiple parts of the decision boundary of the target model uncertainly. Moreover, using  multiple reference samples across multiple classes for random augmentation will allow coverage of a wider region of the decision boundary. Consequently, the region of the decision boundary that needs to be fixed by adversarial training is uncertain and covers a wide area of the decision boundary. Further, the misclassification behavior caused by encysted samples generated by our methodologies is also high in uncertainty, generally among multiple classes. Namely, such encysted samples are hard to handle for the capability of the protected model. For example, given a reference sample and selected latent code $z$, the reconstructed samples $x'$ and $x''$, by adding 0.01 and 0.02 perturbations to $z$ respectively, may have different predicted labels from the protected model. $x'$ and $x''$ also share high similarity in the pixel space. As illustrated in Figure~\ref{fig_aar}, three reconstructed images with similar latent perturbations (also visually similar) have totally different predictions. 

For adversarial training conducted by the adaptive attacker, it is reasonable to treat our encysted samples as noisy samples and exclude them from the learning process. Due to the unpredictability, the labels of these encysted samples are usually distributed among multiple classes, which is not imbalanced, resulting in weight loss ineffectiveness. 

Why smoothness is important in the public verification scenario?
\textit{(i)} The dishonest cloud may implement commonly adopted anomaly detection approaches to identify the queried verification samples using their abnormal patterns in pixel space. This is the reason why existing fingerprinting approaches can be defeated, because fingerprinting patterns, such as the trigger pattern and pixel-wise perturbation pattern, can be easily identified. The fingerprinting design in pixel space and detection approaches are both operating in pixel space, one side always beats the other, as a Min-Max zero-sum game. 
Therefore, powerful attacks could always be designed to undermine the pixel-wise fingerprinting design. 
However, our fingerprinting design is conducted in the latent space, while guaranteeing smoothness in the pixel space as well. 
The reconstruction error via an autoencoder has been demonstrated to be efficient to detect pixel-wise perturbation, detecting more than 90\% pixel perturbation-based verification samples in Section 5.8. However, such detection is inefficient to identify our latent perturbation (2.3\% with the cost of a high false-positive rate), as our encysted sample is generated from the autoencoder with reconstruction error as one regularization term. 
Therefore, our approach is feasible to achieve an indistinguishable verification query from the normal business query, so as to defeat the dishonest could. 
\textit{(ii)} Additionally, we consider extreme adversaries who are capable of collecting a set of publicly available verification samples for all classes. Then, they use these samples to perform adversarial training in order to defeat the verification service.
As we demonstrate in Section~5.8, such adaptive attacks can bypass more than 80\% of verification samples that are designed in pixel space, using 1,000 publicly released verification samples for adversarial training. In other words, the adaptive attacker has successfully learned the distribution of fingerprinting patterns in the pixel space. Contrary to this, our approach remains unchanged after applying adversarial training with the same amount of publicly available verification samples. One reason is that the smoothness in the pixel space restricts the ability to learn fingerprinting patterns from publicly released verification samples.

\section{Cross-validation}\label{sec_cross_validation}
In this work, cross-validation across a model's decision boundary is conducted on the variations from retraining the same model structure on the same dataset with different initialization, producing fingerprints accordingly. Then the fingerprints are randomly selected \textit{w.r.t.} retraining group to perform verification of integrity on each retraining model, resulting in 100\% accuracy with 5 verification samples for all cases. 
It is worth mentioning that the model's quality check and the integrity check are two different businesses. The purpose of the quality check is to determine the quality of the model, including its vulnerability and fragility from the development view. On the contrary,  this work focuses primarily on the integrity check, which verifies whether the queried model is the certified model from the consistency view or if it is different from the model used to produce fingerprinting at the start.

\end{document}